\documentclass[aps,prd,superscriptaddress,nofootinbib,amsfonts,amssymb,amsmath,notitlepage,twocolumn]{revtex4-1}
\usepackage[utf8x]{inputenc}
\usepackage{microtype} 
\usepackage{graphicx}
\usepackage{float}
\usepackage{wrapfig}
\usepackage{bm}
\usepackage{color}
\usepackage[usenames,dvipsnames]{xcolor}
\usepackage{comment}
\usepackage{amstext, slashed}
\usepackage{mathtools}
\usepackage[normalem]{ulem}
\usepackage{hyperref}
\usepackage{mhchem}
\usepackage{dcolumn}
\usepackage[noabbrev]{cleveref}
\usepackage{booktabs}
\usepackage{dsfont}
\hypersetup{
colorlinks=true,
citecolor=blue,
citebordercolor=red,
linktoc=all,
linkcolor=blue,
urlcolor=blue
}
\allowdisplaybreaks

\catcode`\@=11
\newcommand{\lsim}{\lesssim}

\def\@versim#1#2{\vcenter{\offinterlineskip
\ialign{$\m@th#1\hfil##\hfil$\crcr#2\crcr\sim\crcr } }}
\catcode`\@=12

\newcommand{\p}{\partial}

\newcommand{\al}[1]{\begin{align}#1\end{align}}
\newcommand{\bp}{\begin{pmatrix}}
\newcommand{\ep}{\end{pmatrix}}
\newcommand{\nn}{\nonumber\\}

\newcommand{\df}{\text{d}}

\newcommand{\bs}[1]{\boldsymbol}

\newcommand{\fn}[1]{\!\left(#1\right)}

\graphicspath{{./figs/}}
\newbox{\ORCIDicon}
\sbox{\ORCIDicon}{\large
                  \includegraphics[width=0.8em]{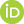}}

\begin{document}
\begin{flushright}
YITP-22-130
\end{flushright}
\vspace{-1cm}
\title{Scaling solutions for asymptotically free quantum gravity}

\author{Saswato \surname{Sen}\,\href{https://orcid.org/0000-0003-2470-3483}{\usebox{\ORCIDicon}}}
\email{saswato.sen@oist.jp}
\affiliation{Okinawa Institute of Science and Technology Graduate University 1919-1 Tancha, Onna, Kunigami, Okinawa, Japan 904-0412}

\author{Christof \surname{Wetterich}\,\href{https://orcid.org/0000-0002-2563-9826}{\usebox{\ORCIDicon}}}
\email{c.wetterich@thphys.uni-heidelberg.de}
\affiliation{Institut f\"ur Theoretische Physik, Universit\"at Heidelberg, Philosophenweg 16, 69120 Heidelberg, Germany}

\author{Masatoshi \surname{Yamada}\,\href{https://orcid.org/0000-0002-1013-8631}{\usebox{\ORCIDicon}}\, }
\email{m.yamada@thphys.uni-heidelberg.de}
\affiliation{Institut f\"ur Theoretische Physik, Universit\"at Heidelberg, Philosophenweg 16, 69120 Heidelberg, Germany}
\affiliation{Center for Theoretical Physics and College of Physics, Jilin University, Changchun 130012, China}

\begin{abstract}
We compute scaling solutions of functional flow equations for quantum gravity in a general truncation with up to four derivatives of the metric. They connect the asymptotically free ultraviolet fixed point, which is accessible to perturbation theory, to the non-perturbative infrared region. The existence of such scaling solutions is necessary for a renormalizable quantum field theory of gravity. If the proposed scaling solution is confirmed beyond our approximations asymptotic freedom is a viable alternative to asymptotic safety for quantum gravity.
\end{abstract}

\maketitle

\section{Introduction}
Long ago it was established that quantum gravity with an action containing up to four derivatives is perturbatively renormalizable~\cite{Stelle:1976gc} and asymptotically free~\cite{Fradkin:1978yf,Fradkin:1981hx,Fradkin:1981iu,Julve:1978xn,Avramidi:1985ki,Avramidi:1986mj,Antoniadis:1992xu,deBerredo-Peixoto:2003jda,deBerredoPeixoto:2004if}. Including in the action also a term linear in the curvature scalar $R$ leads, however, to tachyonic and ghost instabilities. Recent perturbative investigations~\cite{Salvio:2014soa,Salvio:2017qkx} point to the need of understanding the flow of couplings outside the perturbative region. Indeed, the question of instabilities concerns the behavior of the full propagator for the graviton and other physical modes. These propagators are given by the inverse of the second functional derivative of the quantum effective action, rather than the classical action. Inclusion of the effects of quantum fluctuations is crucial for settling the issue of potential instabilities. 

A polynomial expansion of the inverse propagator in powers of momentum necessarily leads to ghosts and tachyons if it is cut at any finite order beyond second order. These instabilities may be an artifact of the truncation~\cite{Donoghue:2019fcb,Platania:2020knd,Platania:2022gtt,Wetterich:2019qzx}, as revealed by examples for the full momentum dependence of acceptable propagators~\cite{Christiansen:2014raa,Christiansen:2015rva,Denz:2016qks,Bosma:2019aiu,Knorr:2019atm,Wetterich:2019qzx,Bonanno:2021squ,Knorr:2021niv,Wetterich:2021ywr,Fehre:2021eob}. 
The difference between the quantum effective action and the classical or microscopic action is due to quantum fluctuations.
These quantum fluctuations are responsible for running couplings. Since for asymptotically free quantum gravity the flow of the couplings necessarily quits the perturbative domain, only a non-perturbative study can answer the question if asymptotically free quantum gravity is an acceptable renormalizable quantum field theory or not.   

Functional flow equations~\cite{Wetterich:1992yh,Reuter:1993kw,Reuter:1996cp} have opened the door for non-perturbative investigations of the effects of the fluctuations of the metric. They have revealed the possible existence of a non-perturbative ``Reuter fixed point". If this fixed point is realized in the ultraviolet (UV), quantum gravity is non-perturbatively renormalizable or asymptotically safe~\cite{Hawking:1979ig,Reuter:1996cp,Souma:1999at,Niedermaier:2006wt,Niedermaier:2006ns,Percacci:2007sz,Reuter:2012id,Codello:2008vh,Eichhorn:2017egq,Percacci:2017fkn,Eichhorn:2018yfc,Reuter:2019byg,Bonanno:2020bil,Reichert:2020mja}. This fixed point has been found for  a large variety of truncations of the effective average action for pure gravity~\cite{Lauscher:2002sq,Codello:2006in,%
Falls:2013bv,Falls:2014tra,Falls:2017lst,Falls:2018ylp,Kluth:2020bdv,Kluth:2022vnq,%
Dietz:2012ic,Dietz:2013sba,%
deBrito:2018jxt,Ohta:2013uca,Ohta:2015zwa,%
Benedetti:2009rx,Benedetti:2009gn,Benedetti:2010nr,Groh:2011vn,%
Manrique:2010am,%
Donkin:2012ud,%
Christiansen:2012rx,Christiansen:2016sjn,Christiansen:2017bsy,%
Eichhorn:2018akn,Eichhorn:2018ydy,%
Codello:2013fpa,Demmel:2014hla,Biemans:2016rvp,Gies:2016con,deBrito:2020rwu,deBrito:2020xhy,deBrito:2021pmw,%
Gonzalez-Martin:2017gza,Baldazzi:2021orb,Baldazzi:2021fye,deBrito:2022vbr,Falls:2020qhj,Knorr:2021slg,%
Mitchell:2021qjr,Morris:2022btf}
and for gravity-matter systems~\cite{Dona:2013qba,Dona:2015tnf,Percacci:2015wwa,%
Oda:2015sma,Hamada:2017rvn,%
Labus:2015ska,%
Eichhorn:2016esv,Eichhorn:2015bna,%
Christiansen:2017cxa,Meibohm:2016mkp,%
Biemans:2017zca,%
deBrito:2019epw,Eichhorn:2018nda,%
Alkofer:2018fxj,Alkofer:2018baq,%
Burger:2019upn,%
deBrito:2019umw,deBrito:2020dta,Eichhorn:2020kca,Eichhorn:2020sbo,Eichhorn:2021tsx,%
Ohta:2021bkc,Laporte:2021kyp,Knorr:2022ilz,Hamada:2020mug,Eichhorn:2022vgp,Wetterich:2022bha,%
Pastor-Gutierrez:2022nki}.
The asymptotic safety scenario provides strong predictivity for matter interactions. This fact has been applied for the standard model~\cite{Shaposhnikov:2009pv,Eichhorn:2017eht,Eichhorn:2017ylw,Eichhorn:2018whv,Alkofer:2020vtb,Eichhorn:2017muy,Harst:2011zx,Christiansen:2017gtg,Eichhorn:2017lry} and its extensions~\cite{Eichhorn:2019dhg,Eichhorn:2017als,Reichert:2019car,Hamada:2020vnf,%
Kowalska:2020zve,Kowalska:2020gie,Kowalska:2022ypk,Chikkaballi:2022urc,%
Boos:2022jvc,Boos:2022pyq,deBrito:2021akp}.
Recently, the flow equations have been derived for the most general effective action for the metric with diffeomorphism invariance and up to four derivatives, covering fully the perturbative and non-perturbative range of all couplings~\cite{Sen:2021ffc}. Besides the asymptotically safe fixed point, these flow equations also allow for a non-perturbative investigation of asymptotically free quantum gravity. This is the topic of the present paper.   

Within functional flow equations the UV completion of quantum gravity does not only require a fixed point behavior of a small finite number of dimensionless couplings. Whole coupling functions of fields and momenta have to admit a scale invariant form, such that the model can be extrapolated to infinitely high momenta or arbitrarily small distances. In particular, in the presence of scalar fields functions as the effective scalar potential or the field dependent curvature coefficient (``squared effective Planck mass") have to take a scale invariant form. In the present paper, we demonstrate within our truncation that such ``scaling solutions" connecting a perturbative range for small scalar fields to a non-perturbative range of large scalar fields indeed do exist. 

For scaling solutions the coupling functions depend only on the dimensionless ratio $\tilde\chi=\chi/k$ where $\chi$ is a scalar field and $k$ the renormalization scale. Computing scaling functions, as the dimensionless effective potential $u=U/k^4$, as functions of $\tilde\chi$ permits an interpolation between the UV-region for $k\to \infty$ or $\tilde\chi\to 0$ to the IR-region for $k\to 0$ or $\tilde\chi\to \infty$. We observe that the $\tilde\chi$-dependence of coupling functions according to the scaling solution is directly related to the $k$-dependence of single couplings defined at fixed $\chi=0$ according to the more general solution of the flow equation. Similarly, the same scaling solutions can be used to describe running gravitational couplings in the absence of a scalar field. Our results for the scaling solution therefore translate directly to results of the flow away from the UV-fixed point for a finite number of couplings. In our truncation this concerns four couplings, namely the flowing Planck mass and cosmological constant as well as the coupling $C$ multiplying the squared curvature scalar and $D$ multiplying the squared Weyl tensor.

Computing the full scaling functions as $u(\tilde\chi)$ offers, however, substantial additional insight beyond the flow of these four couplings. First of all, scalar fields exist in any realistic model of particle physics coupled to quantum gravity. The scalar field can be the Higgs scalar or the inflaton or cosmon for early or late cosmology. The scaling function $u(\tilde\chi)$ yields the fixed point values for all couplings that are defined by some expansion of $u(\tilde\chi)$, as scalar mass terms, quartic couplings or higher order couplings. For an UV complete theory all these couplings have to assume fixed point values in the UV.

Second, for cosmology one is typically interested in the shape of scalar potentials for a large range of fields, not only in a polynomial expansion for small fields. The understanding of the behavior for large fields becomes necessary if the field value covers a substantial range during the cosmic evolution, as for the example of inflation. The scaling form of $u(\tilde\chi)$ yields already an overall functional shape. This may be used as a starting point in the UV from which the flow may depart for  decreasing $k$ due to the presence of relevant parameters.  We can associate the first substantial departure from the scaling solution to a mass scale $k_c$. If $k_c$ is low enough, the scaling solution governs a very large part of the flow for all $k$ larger than $k_c$.

Third, it is possible that the scaling solution itself describes our world. Such a ``fundamental scale invariance"~\cite{Wetterich:2020cxq} is a highly predictive scheme since relevant parameters at the UV-fixed point play no role. For this type of model the implicit dependence on the renormalization scale $k$ can be removed by switching to scale invariant fields or by an appropriate Weyl scaling of the metric.  

Within our approximation of the most general effective action with up to four derivatives of the metric we find that scaling solutions connecting the asymptotically free UV-fixed point to the non-perturbative IR-region indeed do exist. In the vicinity of the UV-fixed point the couplings $C^{-1}$ and $D^{-1}$ are small. For any given fixed value of the scalar field $\chi$ the dependence on the renormalization scale $k$ follows the perturbative running~\cite{Fradkin:1978yf,Fradkin:1981hx,Fradkin:1981iu,Julve:1978xn,Avramidi:1985ki,Avramidi:1986mj,Antoniadis:1992xu,deBerredo-Peixoto:2003jda,deBerredoPeixoto:2004if}. Both $C^{-1}$ and $D^{-1}$ reach zero for $k\to \infty$, in accordance with asymptotic freedom. This includes the running of $C^{-1}(0)$ and $D^{-1}(0)$ or corresponding couplings in the absence of a scalar field. The logarithmic running of $C^{-1}$ and $D^{-1}$ is very slow, but drives these couplings finally outside the perturbative domain as $k$ decreases. For the non-perturbative region of large $C^{-1}$ or $D^{-1}$ the slow running continues and will be described more quantitatively below. We do not observe any dramatic change or qualitative crossover in the flow of these couplings.
 
\begin{figure}
\includegraphics[width=\columnwidth]{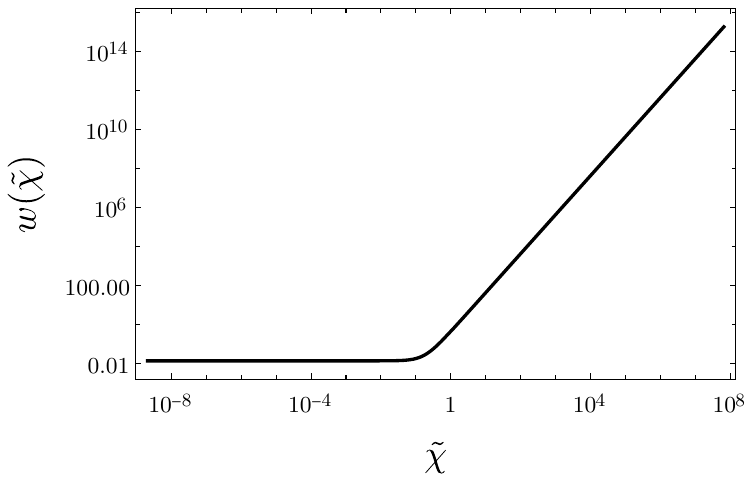}
\caption{
Scaling function $w(\tilde\chi)$ for the coefficient of the curvature scalar (dimensionless squared effective Planck mass). While $w(\tilde\chi)$ becomes almost constant for small $\tilde\chi=\chi/k$, it increases $\sim \tilde\chi^{2}$ for large $\tilde\chi$. For large scalar fields $\chi$ the effective Planck mass is proportional to $\chi$.
}
\label{fig: w} 
\end{figure}

\begin{figure}
\includegraphics[width=\columnwidth]{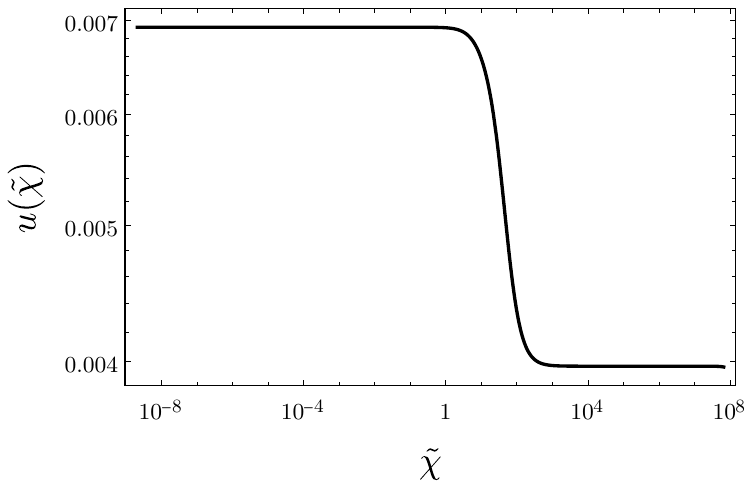}
\caption{
Scaling solution for the dimensionless effective scalar potential $u(\tilde\chi)$ for pure gravity coupled to a scalar field. The potential is almost flat, with a moderate switch between two plateaus at the value of $\tilde\chi=\chi/k$ characteristic for the crossover. 
}
\label{fig: u} 
\end{figure}

 A qualitative crossover is observed in the dimensionless curvature coefficient $w(\tilde\chi)$, which corresponds to a term $\sim wk^2 R$ in the effective action, with $R$ the curvature scalar. For pure gravity with a scalar field the corresponding scaling function is shown in Fig.~\ref{fig: w}.
We observe for large $\tilde\chi$ the behavior $w\sim \tilde\chi^2$, which implies that the effective Planck mass is proportional to the scalar field $\chi$. For small $\tilde\chi$ the scaling function $w(\tilde\chi)$ slowly approaches a fixed point value $w_0$ for $\tilde\chi\to 0$ or $k\to \infty$. The qualitative change in the behavior of $w(\tilde\chi)$ indicates the transition from the UV-region for small $\tilde\chi$ ($k$ above the effective Planck mass) to the IR-region for large $\tilde\chi$ ($k$ below the effective Planck mass). We also observe a mild crossover in the dimensionless scalar potential $u(\tilde\chi)$, as shown in Fig.~\ref{fig: u}.
The potential is almost flat. Its shape is rather different from a polynomial form. The change between the two flat regions at the crossover scale is rather moderate. We will find a similar behavior for models with a different content of particles.

We organize our discussion of the scaling solution in several steps: In \Cref{sec: Effective action for higher derivative gravity} we describe our truncation for the effective action for the metric and a scalar field. \Cref{sec: Flow equations} describes the flow equations and the associated differential equations which define the scaling solutions. \Cref{sec: Fixed points} turns to the fixed points. They are the endpoints of the scaling solutions in the ultraviolet and infrared limits. In \Cref{sec: Scaling solutions} we present the main result of this paper, namely the scaling solution or critical trajectory that links the asymptotically free fixed point to the infrared fixed point. The existence of this scaling solution is a condition for the renormalizability of asymptotically free quantum gravity. \Cref{sec: Various truncations} displays various truncations with a smaller number of coupling functions. They give a rough idea about the robustness of our results. Conclusions are presented in \Cref{sec: Conclusions}.

\section{Effective action for higher derivative gravity}
\label{sec: Effective action for higher derivative gravity}
Our ansatz for the truncated effective average action is 
\al{
\Gamma_k=  \Gamma_k^\text{gravity}+\Gamma_k^\text{matter}\,.
\label{starting effective action}
}
For the gravity part, we consider the following truncated effective action in the Weyl basis,
\al{
\label{Eq: effective action for gravity part}
  \Gamma_k^\text{gravity}
  &=\int \df^4 x\, \sqrt{g}
  \Bigg[U\fn{\rho}  -\frac{F\fn{\rho}}{2} R \nn
&\quad  - \frac{C\fn{\rho}}{2}R^2 +   \frac{D\fn{\rho}}{2} C_{\mu\nu\rho\sigma}C^{\mu\nu\rho\sigma}   \Bigg]\,,
}
where $R$ is the curvature scalar and $C_{\mu\nu\rho\sigma}$ is the Weyl tensor, whose squared form is given by $C_{\mu\nu\rho\sigma}C^{\mu\nu\rho\sigma}=R_{\mu\nu\rho\sigma}^2-2R_{\mu\nu}^2+1/3R^2$. 
The coefficients $U\fn{\rho}$, $F\fn{\rho}$, $C\fn{\rho}$ and $D\fn{\rho}$ are functions of real scalar fields $\chi^a$, $\rho=\chi^a\chi^a/2$, or complex scalar fields $\chi_a$, $\rho=\chi_a^\dagger \chi_a$. 
One can expand these coefficients into a polynomial of $\rho$:
\al{
U(\rho)&= V + m^2 \rho +\frac{\lambda}{2}\rho^2 +\cdots\,,\\[1ex]
F(\rho)&= M_0^2 + \xi \rho + \cdots\,, \\[1ex]
C(\rho)&= C_0 + C_1 \rho +\cdots \,,\\[1ex]
D(\rho)&= D_0 + D_1 \rho +\cdots \,,
}
where $V$ is the cosmological constant; $m^2$ is the scalar mass parameter; $\lambda$ is the quartic coupling; $M_0^2$ is the Planck mass squared at zero scalar field; and $\xi$ is the non-minimal coupling between the scalar field and the curvature scalar and plays a crucial role for the realization of the Higgs inflation.
For the effective average action the functions depend, in addition, on the renormalization scale $k$.

We have not displayed in Eq.~\eqref{Eq: effective action for gravity part} the terms $\Gamma_\text{gf}$ and $\Gamma_\text{gh}$ which encode the gauge fixing and the ghost action for diffeomorphisms, see below. We also do not display here the Gauss-Bonnet term which would be a topological invariant for a constant coupling function. It does not influence the flow of the other couplings. We will discuss this term briefly in \Cref{app: sec: Gauss-Bonnet term}. Furthermore, we have omitted the scalar kinetic term. In our truncation for the flow equations we only consider scalars with canonical kinetic terms and neglect mixings of scalar fluctuations with the scalar metric fluctuation.

We employ a ``physical gauge fixing" which ensures the projection on physical fluctuations for the gauge invariant flow equations~\cite{Wetterich:2016ewc,Wetterich:2017aoy,Wetterich:2017aoy,Pawlowski:2018ixd,Wetterich:2019zdo}. The corresponding gauge fixing and ghost terms are given by
\al{
  \Gamma_\text{gf} &= 
    \frac{1}{2\alpha}\int \df^4x \sqrt{ \bar{g} } \, \bar g^{\mu\nu} \bar D^\alpha h_{\alpha\mu} \bar D^\beta h_{\beta\nu} \,,\label{standard gauge fixing}\\[2ex]
  \Gamma_\text{gh} &= \int \df^4x \sqrt{ \bar{g} } \,\bar C_\mu \left[ \bar g^{\mu\nu}\bar\Delta_V - \bar D^\mu \bar D^\nu - \bar R^{\mu\nu}\right] C_\nu\,,
  \label{standard ghost action}
}
in the limit $\alpha\to0$. Here $C_\mu$ and $\bar C_\mu$ are the ghost and anti-ghost fields, respectively and $\bar \Delta_V=-\bar D^2$ is the Laplacian acting on a vector field.
(These actions correspond to $\beta=-1$ in the standard forms of the gauge fixing for the metric. )

For the matter part $\Gamma_k^\text{matter}$, we consider $N_S$ scalar bosons, $N_V$ vector bosons and $N_F$ Weyl fermions as free massless particles.
In this approximation the gauge and Yukawa couplings are assumed to vanish. We take canonical kinetic terms and neglect a possible mixing of scalars with the physical scalar fluctuation of the metric.

\section{Flow equations}
\label{sec: Flow equations}
Using the functional renormalization group~\cite{Wetterich:1992yh,Tetradis:1992qt,Morris:1993qb,Tetradis:1993ts,Reuter:1993kw,Ellwanger:1993mw,%
Morris:1998da,Berges:2000ew,Aoki:2000wm,Bagnuls:2000ae,%
  Polonyi:2001se,Pawlowski:2005xe,Gies:2006wv,Delamotte:2007pf,%
  Rosten:2010vm,Kopietz:2010zz,Braun:2011pp,Dupuis:2020fhh} with the expansion of the metric field into a background field and a fluctuation one, $g_{\mu\nu}=\bar g_{\mu\nu}+h_{\mu\nu}$, the flow of the effective action~\eqref{starting effective action} is derived. The flow equation for the system is schematically written as ($\p_t = k\p_k$)
\al{
\p_t\Gamma_k
&=\left[N_S\pi^{(S)}_k
+N_V\left(\pi_k^{(V)}-\delta_k^{(V)}\right)
+N_F\pi^{(F)}_k\right]\nn
&\quad
+\pi_k^{(t)}+\pi_k^{(\sigma)}-\delta_k^{(g)}\,.
\label{eq: general flow equation}
}
The flow generators for scalars, gauge bosons and Weyl fermions are denoted by $\pi^{(S)}_k$, $\zeta_k^{(V)}=\pi_k^{(V)}-\delta_k^{(V)}$ and $\pi^{(F)}_k$, respectively.
The last three terms are contributions from metric fluctuations, where $\pi^{(t)}$ and $\pi^{(\sigma)}$ denote contributions from the transverse-traceless (TT) spin-2 tensor and the spin-0 scalar mode, respectively, while contributions from the  gauge fluctuations (the longitudinal modes in metric fluctuations and the ghost fields) are included in the universal measure contribution $\delta^{(g)}$.
From Eq.~\eqref{eq: general flow equation}, we obtain the flow equations for each coupling function according to Ref.~\cite{Sen:2021ffc} as
\begin{widetext}
\al{
\p_t u&=\beta_U=2{\tilde \rho}\,\p_{\tilde \rho}u-4u+\frac{1}{32\pi^2}\left(N_S +2N_V-2N_F +M_U\right)\,,
\label{eq: beta function of U}
\\[2ex]
\p_t w&=\beta_F=2{\tilde \rho}\,\p_{\tilde \rho}w-2w - \frac{1}{96\pi^2}\left( N_S - 4N_V + N_F+M_F \right)\,,
\\[2ex]
\p_t C&=\beta_C=2{\tilde \rho}\,\p_{\tilde \rho} C-\frac{1}{576\pi^2}\left( N_S+M_C\right)\,,\\[2ex]
\p_t D&=\beta_D=2{\tilde \rho}\,\p_{\tilde \rho} D+ \frac{1}{960\pi^2}\left( N_S+12N_V+3N_F +M_D \right)\,.
\label{eq: beta function of E}
}
\end{widetext}
For these flow equations the $k$-derivative is taken for fixed 
\al{
\tilde\rho =\frac{\rho}{k^2}\,.
}
The variable change from $\rho$ to $\tilde\rho$ for the field hold fixed for the partial $k$-derivative results in the contributions $\sim \tilde\rho\, \p_{\tilde\rho}$. We further have introduced the dimensionless combinations
\al{
&u =\frac{U}{k^4}\,,&
&w= \frac{F}{2k^2}\,.
}
The flow kernels $M_i$ specify the contributions from metric fluctuations. They depend on 
\al{
&v=u/w\,,& &c=C/w\,,&&d=D/w\,,
}
as well as $w$. 
The flow equations are valid for arbitrary $\tilde\rho$ (no restriction to solutions of field equations). For this reason the field-dependent propagators for the graviton fluctuations and scalar metric fluctuations involve mass terms $m_t^2$ and $m_\sigma^2$. Their dimensionless form is given by
\al{
&\tilde m_t^2=\frac{m_t^2}{k^2}=\frac{D-u}{w}\,,&
&\tilde m_\sigma^2=\frac{m_\sigma^2}{k^2}=\frac{3C-u/4}{w}\,.
\label{eq: mass parameters}
}
The flow equations involve threshold functions which depend on these mass terms.
The explicit forms of $M_i$ are listed in \Cref{app: flow kernels}.

We look for the scaling solutions to the system of differential equations \eqref{eq: beta function of U}--\eqref{eq: beta function of E}. These are solutions which only depend on $\tilde\rho$ and not separately on $k$.
The existence of such a scaling solution is a necessary and sufficient condition for an ultraviolet complete quantum field theory of gravity. A scaling solution requires that the functions $u(\tilde\rho)$, $w(\tilde\rho)$, $C(\tilde\rho)$ and $D(\tilde\rho)$ remain finite for all non-zero and finite $\tilde\rho$. For any non-zero $\rho$ we can then take the UV-limit $k\to \infty$ by following the scaling solution to $\tilde\rho\to 0$. This permits an extrapolation to arbitrarily high momenta or short distances. On the other hand, if a scaling solution does not exist one encounters a singularity at some finite $\tilde\rho$. Since for arbitrarily small non-zero $\rho$ this singularity can be encountered for large $k$ no consistent interpolation from microphysics to macrophysics is possible.

The scaling solution shows at fixed $\tilde\rho$ no $k$-dependence of $u$, $w$, $C$ and $D$.
Setting $\p_t u=\p_t w=\p_t C=\p_t D=0$ in Eqs.~\eqref{eq: beta function of U}--\eqref{eq: beta function of E}, the scaling solutions have to obey the differential equations
\al{
&\p_\tau u(\tilde\chi)=-4u(\tilde\chi)+\frac{1}{32\pi^2}\left(N_S +2N_V-2N_F +M_U\right)\,,
\label{eq: u full}
\\[1ex]
&\p_\tau w(\tilde\chi)=-2w(\tilde\chi) - \frac{1}{96\pi^2}\left( N_S - 4N_V + N_F+M_F\right)\,,
\label{eq: w full}
\\[1ex]
&\p_\tau C(\tilde\chi)=-\frac{1}{576\pi^2}\left( N_S+M_C\right)\,,
\label{eq: C full}
\\[1ex]
&\p_\tau D(\tilde\chi)= \frac{1}{960\pi^2}\left( N_S+12N_V+3N_F +M_D \right)\,.
\label{eq: D full}
}
Here we have defined a dimensionless sliding scale as
\al{
\tau=-\log \sqrt{2\tilde\rho}=-\log |\tilde \chi|\,,
\label{eq: scales tau rho}
}
with $\tilde\chi=\chi/k=\sqrt{2\rho/k^2}$ the dimensionless field value.
For scaling solutions the only dependence on $k$ arises implicitly through the dependence on $\tilde\chi$. 
Our aim is to find solutions for the system of differential equations \eqref{eq: u full}--\eqref{eq: D full} for the whole range $0< \tilde\rho< \infty$ or $-\infty < \tau < \infty$. 

Scaling solutions for the dependence of $u$, $w$, $C$, $D$ on the scalar field $\tilde\chi$ translate directly to flow trajectories in a model of gravity without a scalar field.
If no scalar field is present, we omit the parts involving $\tilde\rho\,\p_{\tilde\rho}$ in the flow equation \eqref{eq: beta function of U}--\eqref{eq: beta function of E}. The truncation for the pure gravity model involves now four $k$-dependent couplings, namely $u_0$, $w_0$, $C_0$ and $D_0$. If the terms $\sim \tilde\rho\,\p_{\tilde\rho}$ are omitted Eqs.~\eqref{eq: beta function of U}--\eqref{eq: beta function of E} are equivalent to Eqs.~\eqref{eq: u full}--\eqref{eq: D full} with $\tau$ replaced by $t=\log k$. Existence of a scaling solution transfers to the existence of a flow trajectory for these couplings from the UV ($k\to \infty$) to the IR ($k\to 0$). We conclude that for models without the scalar field $\chi$ our scaling solutions describe the flow of purely gravitational couplings $u_0(k)$, $w_0(k)$, $C_0(k)$ and $D_0(k)$ away from the fixed point at $k\to \infty$. Existence of a scaling solution implies then the existence of a trajectory for all values of $k$ for these couplings. These possible trajectories for the flow away from the fixed point extend to gravity with a scalar field if the couplings are defined at $\tilde\rho=0$, e.g. $u_0(k)=u(\tilde\rho=0, k)$ etc., provided the terms $\sim \tilde\rho\,\p_{\tilde\rho}$ can be neglected for $\tilde\rho\to 0$.

\section{Fixed points}
\label{sec: Fixed points}
Fixed points play an important role for the understanding of solutions of the flow equations \eqref{eq: u full}--\eqref{eq: D full}. Typical trajectories relate an ultraviolet (UV)-fixed point for $k\to \infty$ to an infrared (IR)-fixed point for $k\to 0$. For the scaling solutions this translates to $\tilde \chi\to 0$ (UV) and $\tilde \chi\to \infty$ (IR).
For $\tilde\chi\to 0$ or $\tilde \chi \to \infty$ the effective particle numbers $N_S$, $N_F$, $N_V$ become independent of $\tilde\chi$. The terms $M_U$, $M_F$, $M_C$ and $M_D$ do not depend explicitly on $\tilde\chi$, only implicitly via their dependence on $v$, $c$ and $d$. Without an explicit dependence on $\tilde\chi$ there is no dependence on $k$ even for a fixed field $\chi$. Fixed points realize exact quantum scale symmetry~\cite{Wetterich:2019qzx}.

Fixed points are solutions of the scaling equations \eqref{eq: u full}--\eqref{eq: D full} with $\p_\tau u=\p_\tau w =\p_\tau c =\p_\tau d=0$. On the one hand, they govern the behavior of scaling solutions for $\tilde\chi\to 0$ and $\tilde\chi \to \infty$. On the other hand, the UV-fixed points also govern the UV-behavior of arbitrary trajectories for the couplings $u(\tilde\rho=0)$, $w(\tilde\rho)=0$, $c(\tilde\rho)=0$ and $d(\tilde\rho)=0$ according to the flow equations \eqref{eq: beta function of U}--\eqref{eq: beta function of E}. If $u$, $w$, $C$ and $D$ are analytic functions of $\tilde\rho$, or at least their $\tilde\rho$-derivative does not diverge $\sim \tilde\rho^{-1}$ or faster,  we can neglect for $\tilde\rho\to0$ the terms $\sim \tilde\rho \,\p_{\tilde\rho}$ in Eqs.~\eqref{eq: beta function of U}--\eqref{eq: beta function of E}, such that $\p_t$ and $\p_\tau$ coincide for $k\to 0$. 
This extends to the UV-behavior of the coupling functions evaluated at arbitrary fixed finite $\rho$, since $k\to \infty$ implies $\tilde\rho\to 0$.
 This holds similarly if we investigate $C^{-1}$ and $D^{-1}$ instead of $C$ and $D$. The independence of the couplings from $\tilde\rho$ for $\tilde\rho\to 0$ is equivalent to the independence of the couplings defined at fixed finite $\rho$ from $k$ in the limit $k\to \infty$. Replacing the coupling functions $u(\tilde\chi)$ etc. by $\tilde\chi$-independent couplings $u(0)$ etc. our investigation of the UV-fixed points for the scaling solution coincides with investigations for fixed points in the $k$-dependence for these particular field-independent couplings. Again, this extends to fixed points for the four couplings in a gravity model without the scalar $\chi$.

We find several types of fixed points. The first type occurs for non-zero and finite $u_*$, $w_*$, $c_*$ and $d_*$. This asymptotically safe fixed points generalize the Reuter fixed point. Second, we obtain asymptotically free fixed points for $(c^{-1})_*=(d^{-1})_*=0$, with non-zero and finite $u_*$ and $w_*$. Third, the IR-fixed points obey $(w^{-1})_*=0$, $c_*=d_*=0$ with finite non-zero $u_*$. The scaling solutions connect a UV-fixed point with an IR-fixed point. For the UV-fixed point one may take either the asymptotically free (AF) or the asymptotically safe (AS) fixed point. 

In the presence of several fixed points the characterization of UV- or IR-fixed points is relative.
The trajectories linking the AF-and AS-fixed point flow from the AF-fixed point to the AS-fixed point as $\tilde\rho$ increases. 
Being exactly on this trajectory the AS-fixed point would be the IR-fixed point. For trajectories very close to this particular trajectory the AS-fixed point is first approached very closely, while the trajectory subsequently turns away from the AS-fixed point and finally reaches the IR-fixed point for $\tilde\rho\to 0$.
The AS-fixed point can therefore be approached as an intermediate approximate fixed point.

We investigate the fixed point values and the corresponding critical exponents in the presence of matter. In a numerical analysis a lot of artifactual fixed points are found. As selection criterions for realistic fixed points, we consider $w_*>0$, $\tilde m_{t*}^2>-1$ and $\tilde m_{\sigma*}^2>-1$. Furthermore, we remove fixed points which yield huge values of the critical exponents $|\theta_i|>100$ because those indicate that the fixed point value is located close to the poles of the propagators of the metric field.

\underline{\it{Asymptotic safety}}:
In \Cref{Table: summery of fixed points}, we summarize the AS-fixed point values for several matter contents. Since $w_*$ is finite one can transfer the quoted values of $C_*$ and $D_*$ to $c_*$ and $d_*$. For the pure gravity case we have reported their numerical values in the previous work~\cite{Sen:2021ffc}. For the standard model particle content (SM) two fixed points survive our rough exclusion criteria. The first with positive $A_*$ generalizes the Reuter fixed point, while the second has negative $A_*$. The quantity $A$ is the gravity induced anomalous dimension of the scalar field, see below. For $A>2$ there is no additional relevant parameter in the scalar sector beyond constant $u$. For $0<A<2$ the scalar mass terms become relevant parameters. Finally, for $A<0$ also the scalar quartic couplings turn to relevant parameters. The fixed points with negative $A$ have therefore a higher number of relevant couplings and are less stable. 

In our truncation we did not find realistic non-trivial AS-fixed points satisfying the selection criterions for the GUT models.
Since such fixed points have been found for truncations with vanishing $C$ and $D$ this may imply that the presence of the four-derivative terms $\sim C$, $D$ is not compatible with an AS-fixed point. It remains possible, however, that our truncation is insufficient for settling this question. It is also possible that on approximate AS-fixed point exists for which the contributions of $C$ and $D$ are subleading and these couplings change slowly.

\begin{table*}
\begin{center}
\begin{tabular}{ l c c c c c c }
\toprule
  \makebox[6cm]{} &   \makebox[2cm]{$u_*$}  &   \makebox[2cm]{$w_*$}  & \makebox[2cm]{$C_*$}  &   \makebox[2cm]{$D_*$}  &   \makebox[2cm]{$A_*$}  \\
\midrule
Pure gravity ($N_S=N_V=N_F=0$) & $0.00028$ & $0.0218$ &  $0.204$ & $-0.0132$ & $0.618$   \\[2ex]
SM ($N_S=4,\,N_V=12,\,N_F=45$) & $-0.0500$ & $0.00321$  & $0.320$ & $-0.0017$  & $0.00638$   \\[2ex]
&  $-0.0580$ & 0.00399 & $-0.0051$ & $-0.023$& $-1.35$  \\[2ex]
SM $+$ a scalar ($N_S=5,\,N_V=12,\,N_F=45$) &  $-0.0492$ & $0.00376$ & $0.318$ & $-0.0020$ & $0.00682$ \\[2ex]
&  $-0.0573$ & $0.00447$ & $-0.00516$ & $-0.0230$ & $-1.36$
\\[2ex]
SU(5) GUT ($N_S=54,\,N_V=24,\,N_F=45$)& --- &  --- &  --- &  ---  &  --- \\[2ex]
SO(10) GUT ($N_S=317,\,N_V=45,\,N_F=48$)&  --- &  --- &  --- &  --- &  --- \\
\bottomrule
\end{tabular}
\caption{
\label{Table: summery of fixed points}
Values for couplings at the AS-fixed point. ``SM" is the abbreviation of the standard model. We also plot the value of the gravity induced scalar anomalous dimension $A_*$ at the AS-fixed point.
}
\end{center}
\end{table*}

The flow away from the fixed point is governed by the critical exponents $\theta$ which correspond to eigenvalues of the ``stability matrix".  This matrix governs the flow of the linear deviations from the fixed point. Positive $\theta$ lead to relevant parameters which result in free couplings, while for negative $\theta$ the associated irrelevant couplings are predicted to take their fixed point values even for the linearized flow away from the fixed point. We plot the critical exponents for the system of flow equations for field-independent $u$, $w$, $C$ and $D$ for the AS-fixed points in \Cref{Table: summery of critical exponents}.

Other relevant parameters may arise in the scalar sector, associated to the dimensionless mass parameter $\tilde m^2_0$ or quartic coupling,
\al{
&\tilde m_0^2 =\frac{\p u}{\p\tilde\rho}(\tilde \rho=0)\,,&
&\tilde \lambda_0 =\frac{\p^2 u}{\p\tilde\rho^2}(\tilde \rho=0)\,.
}
The flow equations for these parameters are found by taking $\tilde\rho$-derivatives of Eq.~\eqref{eq: beta function of U}, evaluated at $\rho=0$.

\begin{table*}
\begin{center}
\begin{tabular}{ l c c c c c }
\toprule
  \makebox[7cm]{} &   \makebox[2cm]{$\theta_1$}  &   \makebox[2cm]{$\theta_2$}  & \makebox[2cm]{$\theta_3$}  &   \makebox[2cm]{$\theta_4$}  \\
\midrule
Pure gravity ($N_S=N_V=N_F=0$) & $3.1$ & $2.4$ &  $10.9$ & $−88.1$  \\[2ex]
SM ($N_S=4,\,N_V=12,\,N_F=45$) & $1.1+0.25i$ & $1.1-0.25i$  & $3.96$ & $-3.24$ \\[2ex]
&  $4.1$ & $2.2$  & $21.2$ & $-3.09$ \\[2ex]
SM $+$ a scalar ($N_S=5,\,N_V=12,\,N_F=45$)& $1.1 + 0.26i$ & $1.1 - 0.26i$ & $3.95$ & $-3.32$  \\[2ex]
&  $4.1$  & $2.2$ & $20.8$ & $-3.13$ \\[2ex]
SU(5) GUT ($N_S=54,\,N_V=24,\,N_F=45$)&  --- &  --- & --- &  --- \\[2ex]
SO(10) GUT ($N_S=317,\,N_V=45,\,N_F=48$)&  --- &  --- &  --- &  ---  \\
\bottomrule
\end{tabular}
\caption{
\label{Table: summery of critical exponents}
Critical exponents at the AS-fixed point for various matter contents. The rows are the same as for \Cref{Table: summery of fixed points}.
}
\end{center}
\end{table*}
From Eq.~\eqref{eq: beta function of U}, critical exponents for the couplings ($g_n=\{V, \tilde m_0^2, \lambda_0,\ldots \}$) are approximately given by
\al{
\theta_{g_n} \simeq  -\frac{\p\beta_U}{\p g_n}\bigg|_{\tilde \rho=0} = (4 -2n ) - A\,.
}
Here $A$ is the metric-induced anomalous dimension
\al{
A&=\frac{1}{32\pi^2}\frac{\p M_U}{\p u} \bigg|_{\tilde\rho\to 0}\nn
&= \frac{1}{96\pi^2w}\left[ \frac{20( 1+\frac{3}{2}d)}{(1 + \tilde m_t^2)^2} + \frac{\frac{9}{20}(1+ 5c)}{(1+\tilde m_\sigma^2)^2} \right]\bigg|_{\tilde\rho\to 0}\,.
\label{eq: metric-fluctuation induced anomalous dimension}
}
The relation \eqref{eq: scales tau rho} is only approximate since the off-diagonal elements in the stability matrix are neglected.

\underline{\it Asymptotic freedom}:
For the asymptotically free fixed point one has
\al{
&C_*^{-1}=0\,,&
&D_*^{-1}=0\,,
\label{eq: asymptotically free fixed point}
}
while the scalar potential or cosmological constant and the Planck mass squared have the following non-trivial fixed points
\al{
&u_*=
\begin{cases}
0.0069  & (\text{Pure gravity})\,, \\[1ex]
-0.0422 & (\text{SM})\,,\\[1ex]
-0.0414 & (\text{SM $+$ a scalar})\,, \\[1ex]
0.0164  & (\text{SU(5) GUT})\,,\\[1ex]
0.2531  & (\text{SO(10) GUT})\,,
\end{cases}
\label{eq: AFF for u}
 \\[3ex]
&w_*=
\begin{cases}
0.0127 & (\text{Pure gravity})\,, \\[1ex]
0.0134 & (\text{SM})\,,\\[1ex]
0.0139 & (\text{SM $+$ a scalar})\,,\\[1ex]
0.0144 & (\text{SU(5) GUT})\,,\\[1ex]
0.1104 & (\text{SO(10) GUT})\,.
\label{eq: AFF for w}
\end{cases}
}
With finite $w_*$ the AF-fixed point also obey $c_*^{-1}=d_*^{-1}=0$. For the GUT-models the AF-fixed points seem to be the only viable UV-fixed points in our truncation.

At the AF-fixed points, the critical exponents are given as canonical scalings:
\al{
&\theta_1=4\,,&
&\theta_2=2\,,&
&\theta_3=0\,,&
&\theta_4=0\,.
}
The gravity induced scalar anomalous dimension $A$ is zero. With $\tilde m_t^2\sim d$, $\tilde m_\sigma^2\sim c$ the two terms in Eq.~\eqref{eq: metric-fluctuation induced anomalous dimension} vanish $\sim d^{-1}$ and $c^{-1}$.
For the asymptotically free fixed points \eqref{eq: asymptotically free fixed point}, \eqref{eq: AFF for u} and \eqref{eq: AFF for w}, the propagators of TT graviton and the physical scalar mode in the metric field behave as
\al{
G_t(q^2) &= \frac{1}{w_*q^2+D_*q^4 -u_*} \sim \frac{1}{q^4}\,,\\
G_\sigma(q^2) &= \frac{1}{w_*q^2+3C_*q^4 -u_*/4} \sim \frac{1}{q^4}\,.
}

\underline{\it Infrared fixed point}:
Finally, for the infrared (IR) fixed point $w= g_N^{-1}\to \infty$, the flow equation for the cosmological constant reads
\al{
\p_t u =4u -\frac{301}{1920\pi^2}+\frac{1}{32\pi^2}\left(N_S +2N_V-2N_F\right)\,.
}
This has a fixed point
\al{
u_*&=\frac{301}{7680\pi^2} + \frac{N_S +2N_V-2N_F}{128\pi^2}\,.
\label{eq: IR fixed point general}
}
In the deep IR for $k\to 0$ only the metric fluctuations, the photon and the cosmon matter, $N_V=1$, $N_S=1$, $N_F=0$, resulting in 
\al{
u_*=0.00635\,.
}
For somewhat larger $k$ the dimensionless potential $u$ may be attracted towards an effective approximate fixed point, corresponding to different numbers of effectively massless particles. Characteristic quantitative values are 
\al{
u_*&\approx 
\begin{cases}
+0.0040 & \text{(Pure gravity)}\,,\\[1ex]
-0.0451 & \text{(SM)}\,,\\[1ex]
-0.0443 & \text{(SM $+$ a scalar)}\,,\\[1ex]
+0.0135 & \text{(SU(5) GUT)}\,,\\[1ex]
+0.2502& \text{(SO(10) GUT)}\,.
\end{cases}
\label{eq: infrared fixed point}
}
The IR-fixed point plays an important role for resolving the cosmological constant problem~\cite{Wetterich:2017ixo,Wetterich:2018qsl}.

\section{Scaling solutions:
Critical flow trajectories from asymptotically free fixed point to infrared fixed point}
\label{sec: Scaling solutions}

The completion of a model of quantum gravity by an UV-fixed point needs not only a fixed point for a small finite number  of couplings. Whole functions of $\tilde\rho$, which are equivalent to infinitely many couplings, need to take fixed values. These are the scaling solutions. We focus on the scaling solutions for the four functions $u(\tilde\rho)$, $w(\tilde\rho)$, $c(\tilde\rho)$ and $d(\tilde\rho)$.  Here $u(\tilde\rho)$ contains all scalar self-interactions at zero momentum, as $\tilde m_0$, $\lambda_0$,, $\p^3 \tilde u/\p \tilde\rho^3(\tilde\rho=0)$ or same other expansion. The function $w(\tilde\rho)$ describes the flowing Planck mass at zero scalar field $w_0=w(\tilde\rho=0)$ or the non-minimal gravitational coupling of the scalar field $\xi_0=(\p w/\p \tilde\rho)(\tilde\rho=0)$ etc.. We do not pay attention here to a further momentum dependence of the couplings. Conceptually, the whole functional $\Gamma$ has to take a scaling form.

\begin{figure}
\includegraphics[width=\columnwidth]{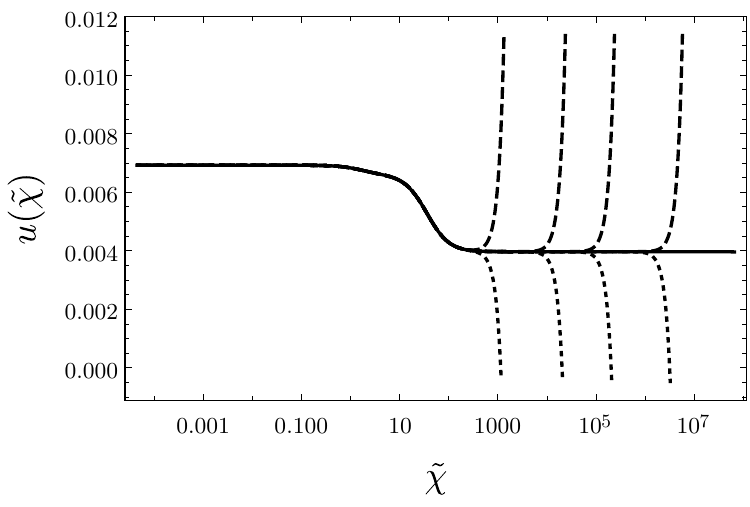}
\caption{
Flow trajectories of $u$ as functions of $\tilde \chi$ with different conditions, for $N_S=1$, $N_F=N_V=0$. 
}
\label{fig: u tuned} 
\end{figure}

We have to find solutions for the system of flow equations  \eqref{eq: u full}--\eqref{eq: D full} for the whole range $0\leq \tilde \rho <\infty$ or $-\infty < \tau < \infty$. The existence of such solutions is much more restricted than the general solution of flow equations. This is demonstrated in Fig.~\ref{fig: u tuned} where the dashed and dotted lines show general flow trajectories, i.e. numerical solutions of Eqs.~\eqref{eq: beta function of U}--\eqref{eq: beta function of E}. Only the solid line corresponds to the scaling solution which extends to the whole range $\tilde\chi\to \infty$. Only for the scaling solution $u(\tilde\chi\to \infty)$ approaches the IR-fixed point \eqref{eq: IR fixed point general} -- in our example for $N_S=1$, $N_V=N_F=0$. The difference in the fixed point values of $u$ for $\tilde\chi\to 0$ and $\tilde\chi\to \infty$ results from the different values for $w$, $c$ and $d$ in the two limits.

\begin{figure}
\includegraphics[width=\columnwidth]{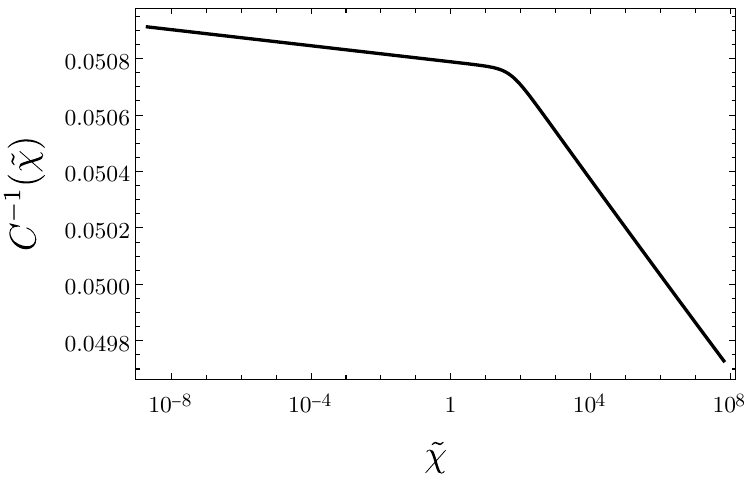}
\hspace{2ex}
\includegraphics[width=\columnwidth]{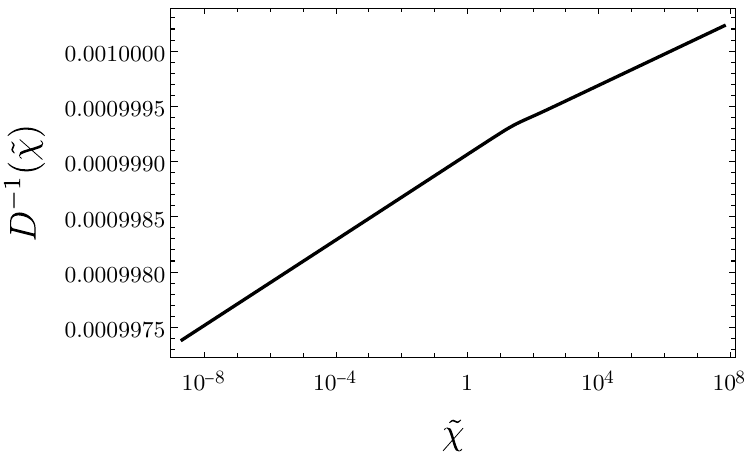}
\caption{
Critical trajectory (scaling solution) of $C^{-1}$ (top) and $D^{-1}$ (bottom).
These trajectories are for $N_S=1$, $N_V=N_F=0$ with initial conditions described in the text.
}
\label{fig: w and u and C and D} 
\end{figure}

\begin{figure}
\includegraphics[width=\columnwidth]{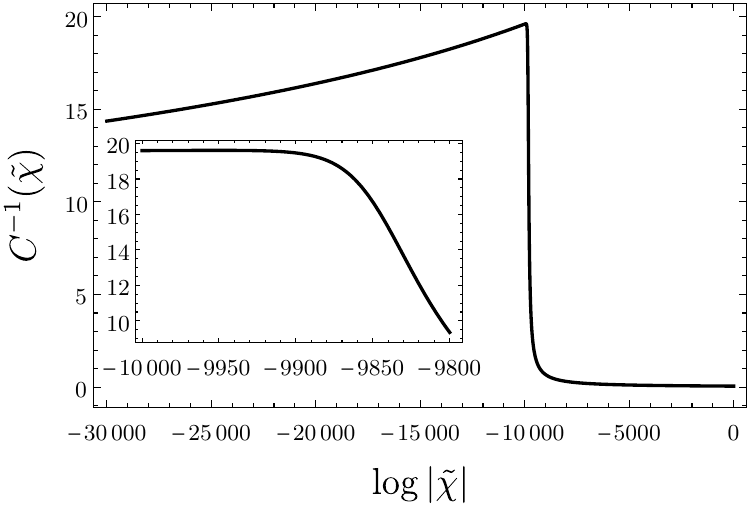}
\caption{
Critical trajectory of $C^{-1}$ as functions of $-\tau=\log|\tilde \chi|$ in a wide range, for $N_S=1$, $N_V=N_F=0$.  We observe the very slow decrease of $C^{-1}$ to the fixed point value zero as $\tilde\chi\to 0$. The inset demonstrates that the peak is smooth.
}
\label{fig: cinv wide} 
\end{figure}

In this note we display numerical scaling solutions which interpolate between the asymptotically free fixed point for $\tilde\chi\to 0$ and the IR-fixed point for $\tilde\chi\to \infty$. In both limits $u$ and $w^{-1}$ approach constants, see Figs.~\ref{fig: w} and \ref{fig: u}. On the other hand, $c^{-1}$ and $d^{-1}$ vanish at the AF-fixed point for $\tilde\chi\to 0$, while $c$ and $d$ vanish for the IR-fixed point. Fig.~\ref{fig: w and u and C and D} shows a critical flow trajectory of the couplings $C^{-1}$ and $D^{-1}$ for pure gravity coupled to a scalar field. We have set the energy scale and initial values such that $w(\tilde\chi)=0.5$, $C^{-1}(\tilde\chi)=0.0508$ and $D^{-1}(\tilde\chi)=9.99065\times 10^{-4}$ at $\tilde\chi=1$. We observe that $D^{-1}$ approaches only very slowly the AF-fixed point $D^{-1}_*=0$. For $C^{-1}$ this approach is even not yet seen in the figure. It occurs at much smaller $\tilde\chi$. We show the wide range of $C^{-1}$ and zoom in on its peak in Fig.~\ref{fig: cinv wide}.

The perturbative range corresponds to small values of $C^{-1}$ and $D^{-1}$, while for the non-perturbative range one has large $C^{-1}$ and $D^{-1}$ or small $C$ and $D$. Fig.~\ref{fig: cinv wide} shows explicitly a non-perturbative range for the coupling $C^{-1}$. We observe that for this particular critical trajectory the flow of $C^{-1}$ first increases slowly as $\tilde\chi$ increases, following the perturbative running near the asymptotically free fixed point. This slow increase continues outside the perturbative range until $C^{-1}$ reaches a maximum. Subsequently, $C^{-1}$ turns back to small values, now decreasing for increasing $\tilde\chi$ due to the presence of the other couplings. There exist other critical trajectories for which $C^{-1}$ remains small for the whole range $\tilde\chi \lsim 1$, similar to the trajectory of $D^{-1}$ shown in Fig.~\ref{fig: w and u and C and D}. 

At the crossover in $w(\tilde\chi)$ from the UV- to the IR-regime (for $\tilde\chi=1$ for our choice of initial values) the slope of the logarithmic running of $C^{-1}$ and $D^{-1}$ changes slightly. The running of these functions does not stop, however. The massless metric fluctuations induce a logarithmic running even for scales $k$ much below the effective Planck mass. For $C^{-1}$ and $D^{-1}$ no decoupling of the metric fluctuations takes place. To the extent that $k$ can be associated with a momentum this implies that the inverse graviton propagator does not have a simple polynomial form. Terms $\sim q^4\log(q^2/\chi^2)$ can have important effects on the stability properties~\cite{Wetterich:2019qzx}.

According to Fig.~\ref{fig: w} the curvature coefficient $w$ approaches for $\tilde\chi\to 0$ the AF-fixed point $w_0$, while it diverges $\sim \tilde\chi^2$ for $\tilde\chi\to \infty$. The bending of $w(\tilde\chi)$ near $\tilde\chi=1$ characterizes the transition from the UV-regime for $\tilde\chi\ll 1$ to the IR-regime for $\tilde\chi\gg 1$. For $\tilde\chi\to \infty$ the scalar potential converges to the IR fixed point \eqref{eq: infrared fixed point}.

\begin{figure}
\includegraphics[width=\columnwidth]{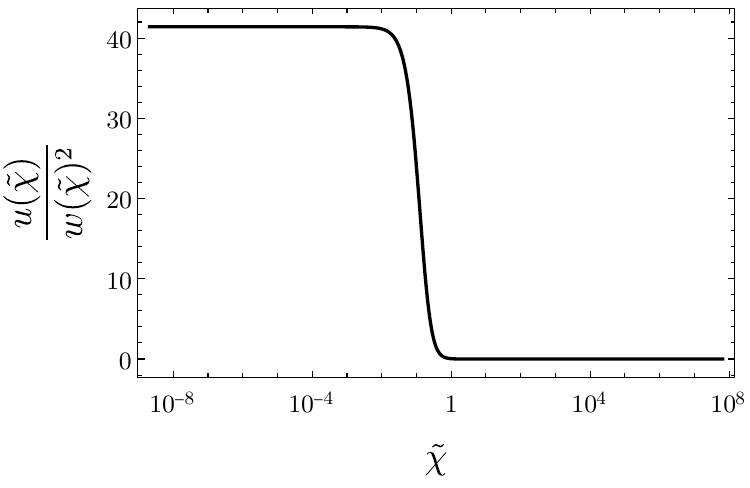}
\caption{
Critical trajectory of $u/w^2$ as function of $\tilde\chi$, for $N_S=1$, $N_V=N_F=0$. 
}
\label{fig: u/w2} 
\end{figure}

\begin{figure}
\includegraphics[width=\columnwidth]{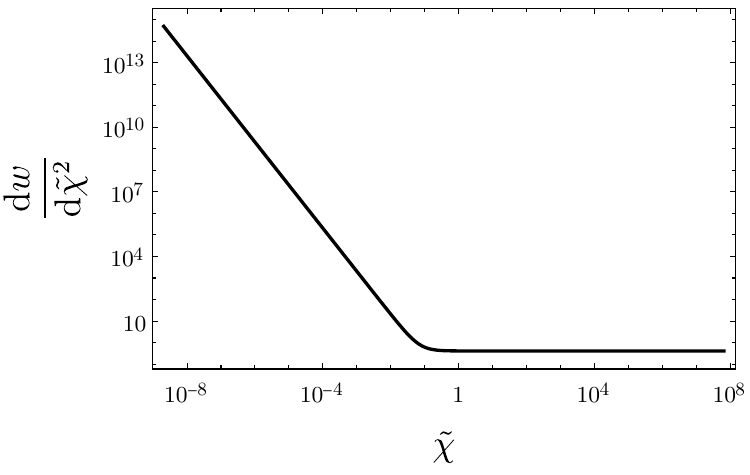}
\caption{
Critical trajectory of $\df w/\df \tilde\chi^2$ as function of $\tilde\chi$, for $N_S=1$, $N_V=N_F=0$. 
}
\label{fig: dwdchi2} 
\end{figure}

The ratio relevant for the observable cosmological constant is $u/w^2$. We plot the scaling solution for this quantity is Fig.~\ref{fig: u/w2}. 
For $\tilde\chi\to 0$ it approaches a positive value $u(0)/w^2(0) =42.8$. For large $\tilde\chi$ the ratio approaches zero rapidly $\sim \chi^{-4}$.  As a consequence, the cosmological constant vanishes asymptotically for $\chi\to \infty$. We also show the non-minimal scalar gravity coupling $\xi=\p w/\p \tilde\rho$ in Fig.~\ref{fig: dwdchi2}. It approaches a constant value $\xi_\infty=0.408$ for $\tilde\chi\to \infty$. On the other side the derivative seems to diverge for $\tilde\chi\to 0$. The strong increase reflects the fact that $w(\tilde\chi)$ flows logarithmically as long as $C^{-1}$ and $D^{-1}$ do not vanish. This flow stops only in the extreme UV-limit $\tilde\chi\to 0$.

\begin{figure*}
\includegraphics[width=\columnwidth]{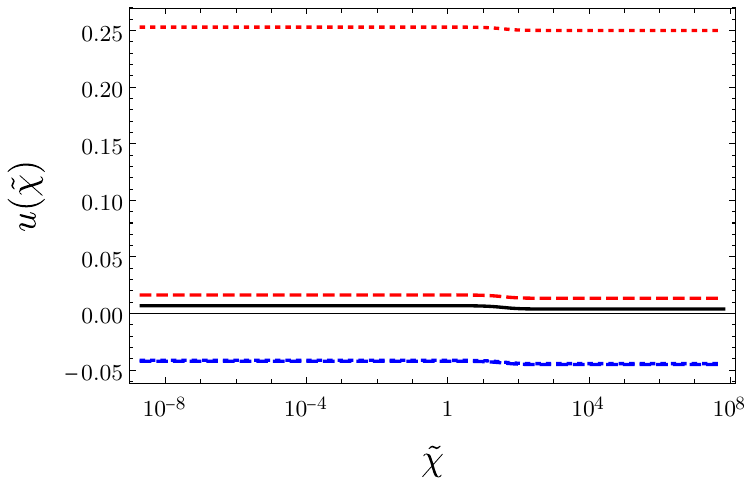}
\hspace{2ex}
\includegraphics[width=\columnwidth]{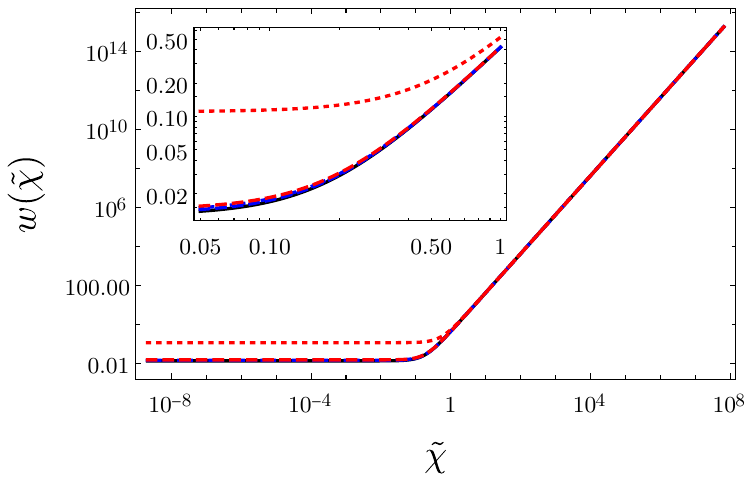}
\hspace{2ex}
\includegraphics[width=\columnwidth]{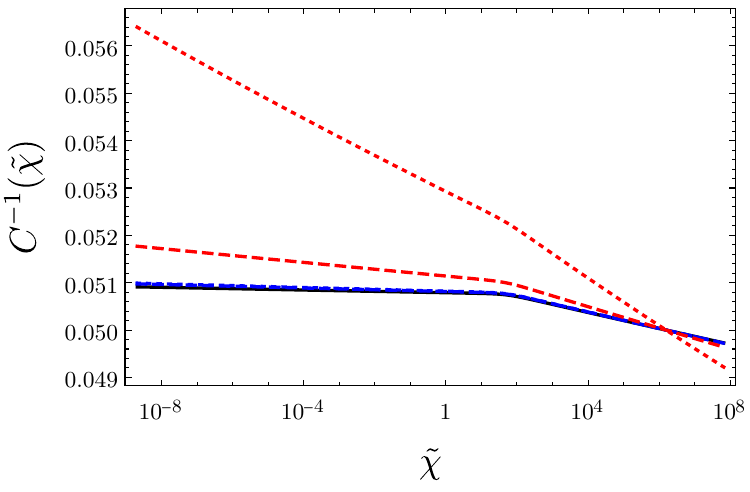}
\hspace{2ex}
\includegraphics[width=\columnwidth]{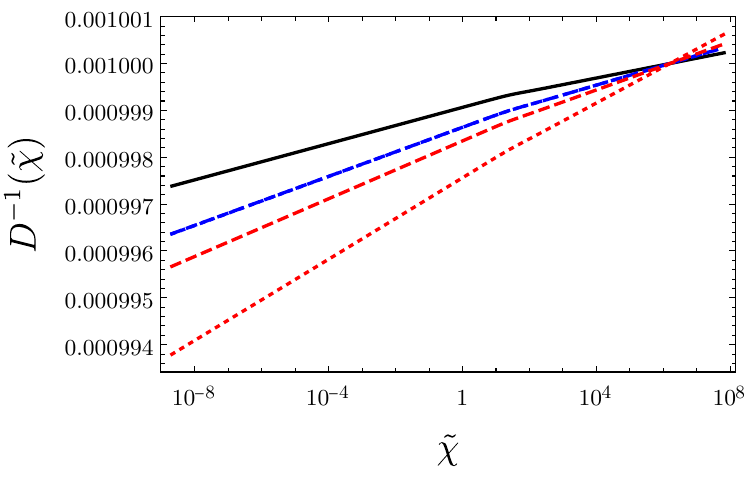}
\hspace{2ex}
\includegraphics[width=\columnwidth]{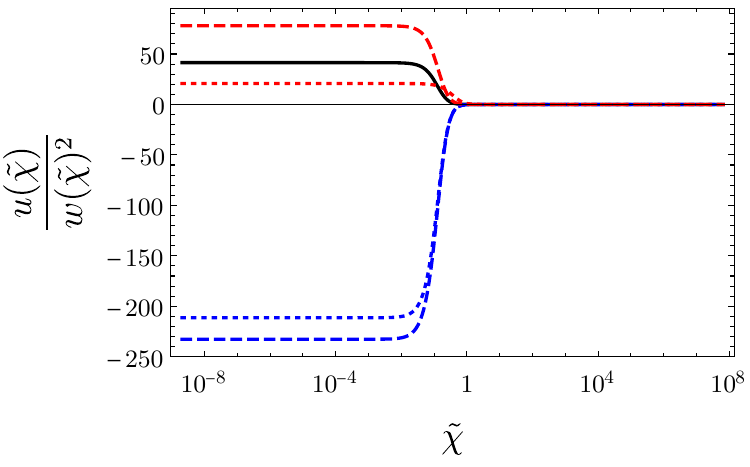}
\hspace{2ex}
\includegraphics[width=\columnwidth]{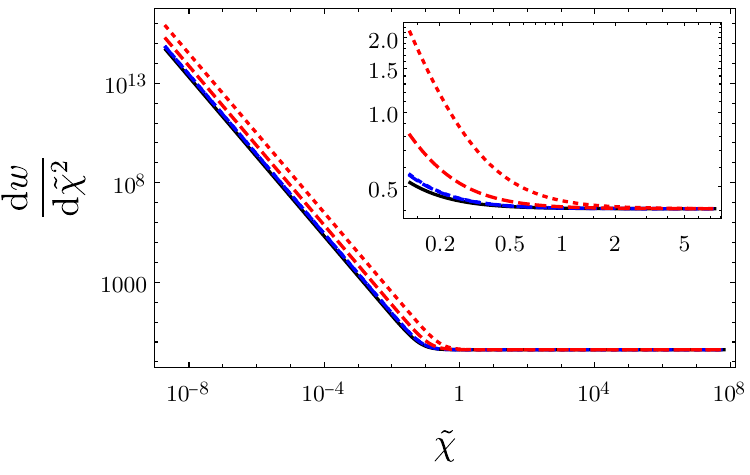}
\caption{
Critical trajectories of $u$ (top-left), $w$ (top-right), $C^{-1}$ (middle-left), $D^{-1}$ (middle-right) $u/w^2$ (bottom-left) and $\df w/\df \tilde\chi^2$ (bottom-right) as functions of $\tilde\chi$ in various models. The solid black line shows the pure gravity case, the blue dashed and dotted lines are the cases of the SM and the SM including a singlet scalar, respectively, and the red dashed and dotted lines are GUTs (SU(5) and SO(10)), respectively.
}
\label{fig: w and u and C and D comp}  
\end{figure*}

The comparison of critical flow trajectories in various models (pure gravity, the SM, the SM plus a singlet scalar, SU(5) and SO(10)) is presented in Fig.~\ref{fig: w and u and C and D comp}. We have set $C(\tilde\chi)^{-1}=0.01$ and $D(\tilde\chi)^{-1}=0.1$ at $\tilde\chi=1.6\times 10^6$. The particle numbers are the ones given in \Cref{Table: summery of fixed points} and \ref{Table: summery of critical exponents}.
We have kept for this plot fixed effective particle numbers $N_S$, $N_F$ and $N_V$. In a more realistic setting some of the particles will decouple effectively in the IR for $\tilde\chi\to \infty$ due to mass terms exceeding $k$ induced by gauge or Yukawa couplings. This will lead to small quantitative changes of the figures.

For all models we observe a very slow decrease of $D^{-1}$ to the fixed point value $D^{-1}(\chi\to 0)=0$, and the decrease of $C^{-1}$ sets in only for much smaller $\tilde\chi$. In view of this very slow running the couplings $C$ and $D$ can be taken as fixed free constants for many purposes. We can indeed find a critical trajectory for a large range of ``initial values" for $C$ and $D$ at some initial $\chi_{\rm ini}$. Within our truncation we find a three parameter family of scaling solutions. They can be parameterized by the values at $\tilde\chi=1$ for the couplings $w$, $C$ and $D$. Thereby the parameter $w(1)$ only sets the scale for $\tilde\chi$. Different $w(1)$ lead to a shift of the value of $\tilde\chi$ where the transition from the UV-region to the IR-region happens. Due to the non-zero values of $C^{-1}$ and $D^{-1}$ or $c^{-1}$ and $d^{-1}$ the coupling $w$ runs logarithmically in the region $\tilde\chi <0.1$
\al{
w=w(\bar\chi) + a (c,d)\log(\tilde\chi^2/\bar\chi^2)\,,
}
with $a\ll w(\bar \chi)$. This leads to the behavior
\al{
\frac{\p w}{\p\tilde\chi^2} = \frac{a}{\tilde\chi^2} 
}
visible in Fig.~\ref{fig: w and u and C and D comp}.
With $a$ involving terms $\sim c^{-1}$ and $d^{-1}$ the very slow running of $w$ towards its fixed point only stops at the fixed point. In practice, this logarithmic running is negligible, as can be seen in the flow of $w(\tilde\chi)$ in Fig.~\ref{fig: w and u and C and D comp}.

The main difference between the various models concerns the value of $u_0=u(0)$.
It is positive for pure gravity and the GUT-models. This influences the shape of the potential in the Einstein frame $U_E=M_{\rm p}^4u/(4w^2)$~\cite{Wetterich:2019rsn,Wetterich:2022brb}. For $u_0>0$ it shows for small $\tilde\chi$ a plateau at positive values which leads for the corresponding cosmology to an inflationary epoch. For increasing $\tilde\chi$ the decrease $U_E(\tilde\chi\to \infty) \to 0$ leads to dynamical dark energy. Such a potential is typical for quintessential inflation~\cite{Spokoiny:1993kt,Peebles:1998qn,Brax:2005uf,Hossain:2014xha,Wetterich:2014gaa,Agarwal:2017wxo,Rubio:2017gty,Geng:2017mic,Dimopoulos:2017zvq,Bettoni:2021qfs,Wetterich:2022brb}. In contrast, for the standard model $u_0$ turns out to be negative. Realistic inflationary cosmology is only possible as a type of Starobinsky inflation~\cite{Starobinsky:1980te} which can be realized for a very large value of $C\approx 10^8$. Critical trajectories with this type of very small $C^{-1}$ (and also $D^{-1}$) exist as part of the family of scaling solutions. This has also been found by a recent dedicated investigation~\cite{Hoshina:2022cws}. 

We have focused the discussion on a real scalar field $\chi$ where $\tilde\rho=\chi^2/(2k^2)$ and $\tau=-\log\sqrt{2\tilde\rho}$. The same flow equations and scaling solutions obtain if $\tilde\rho=\rho/k^2$ is associated with some other quadratic invariant $\rho$ formed from scalar fields. This demonstrates the universal character of the gravitational fluctuation effects. Differences between models with different scalars, or differences in the dependence of the effective potential on different invariants, arise only once the dependence of effective particle numbers on mass thresholds is taken into account. These thresholds depend explicitly on the different invariants.

\section{Various truncations}
\label{sec: Various truncations}
Comparing for $u$ and $w$ the AS- and AF-fixed points one gets the impression of quantitative, but not qualitative differences. This suggests that the inclusion of the couplings $C$ and $D$ does not change the qualitative behavior of the scaling solutions radically. We explore this quantitatively by investigating trajectories where one or both of these couplings are set to zero. Comparison with the full result may give a first idea on the robustness of our results. As an example, we consider ``pure gravity", $N_S=N_F=N_V=0$. 
This allows for a direct comparison with previous work in pure gravity without the scalar field $\chi$. In this case one replaces $\tilde\chi$ by $cM_{\rm p}/k$, where $M_{\rm p}$ is the fixed Planck mass and the constant $c$ is chosen such that $M_{\rm p}^2=F(k=0) =2w(\tilde\chi\to \infty)k^2=\xi_\infty \chi^2/2=(\xi_\infty c^2/2)M_{\rm p}^2$.
For the Einstein-Hilbert truncation and the $f(R)$-type truncation, we present critical flow trajectories from the asymptotically safe fixed point to the IR fixed point.
For the truncation with $C=0$, $D\neq 0$ we look at the trajectory from the AF-fixed point to the IR fixed point.

\subsection{Einstein-Hilbert truncation ($U+R$ system)}
We start with the simplest truncation, i.e. the Einstein-Hilbert truncation where only the scalar potential or cosmological constant and the linear order of the Ricci curvature scalar are taken into account in the effective action ($C=D=0$). In this case, the flow equations read
\al{
\p_\tau u(\tilde\chi)&=-4u(\tilde\chi)+\frac{1}{32\pi^2}M_U(\tilde\chi)\Big|_{\substack{c=0\\[0.5ex]d=0}}\,,
\label{eq: EH: u equation}
\\[1ex]
\p_\tau w(\tilde\chi)&=-2w(\tilde\chi) - \frac{1}{96\pi^2}M_F(\tilde\chi)\Big|_{\substack{c=0\\[0.5ex]d=0}}\,,
\label{eq: EH: w equation}
}
These flow equations admit the UV fixed point,
\al{
&u_*=0.0061\,,&
&w_*=0.022\,.
\label{eq: asymptotically safe fixed point in EH}
}
at which the critical exponents are found to be
\al{
\theta_{1,2}=2.63 \pm 1.39i\,.
}
In addition, there is the IR fixed point \eqref{eq: infrared fixed point}.

\begin{figure*}
\includegraphics[width=\columnwidth]{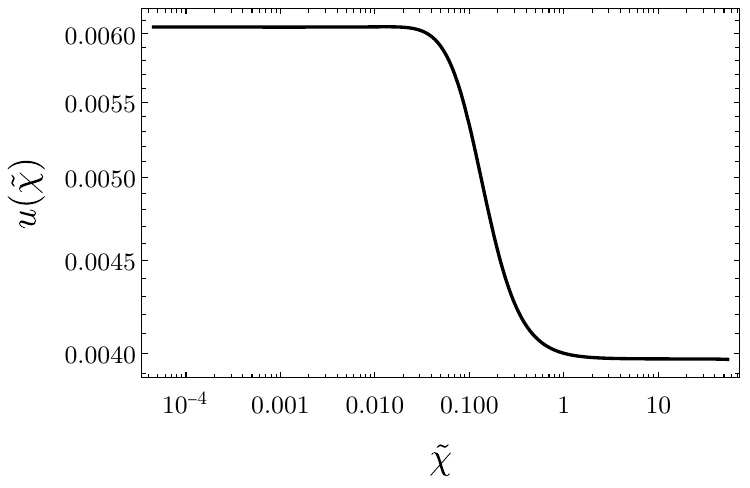}
\hspace{2ex}
\includegraphics[width=\columnwidth]{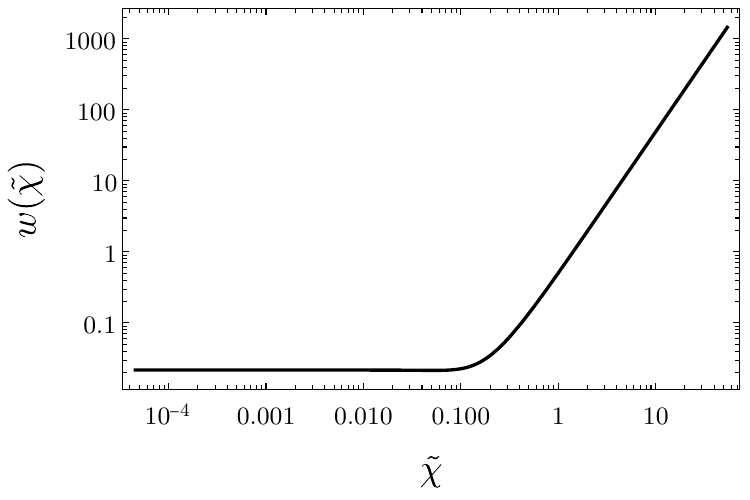}
\includegraphics[width=\columnwidth]{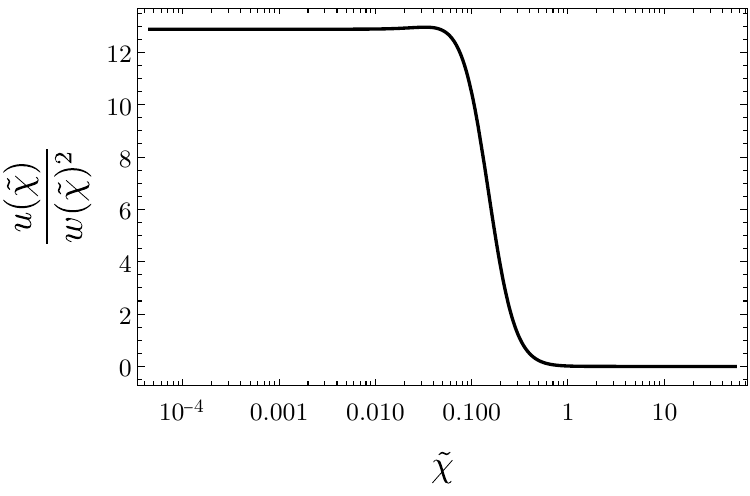}
\caption{
Critical trajectory of $u(\tilde\chi)$ (top-left), $w(\tilde\chi)$ (top-right) and $u(\tilde\chi)/w(\tilde\chi)^2$ (bottom) as functions of $\tilde \chi$  in the Einstein-Hilbert truncation, for $N_S=N_V=N_F=0$. 
}
\label{fig: w and u in EH truncation} 
\end{figure*}

Fig.~\ref{fig: w and u in EH truncation} displays a critical flow trajectory of $u(\tilde\chi)$, $w(\tilde\chi)$ and their ratio $u(\tilde\chi)/w(\tilde\chi)^2$ from the asymptotically safe fixed point \eqref{eq: asymptotically safe fixed point in EH} to the IR fixed point \eqref{eq: infrared fixed point} as a solution to Eqs.~\eqref{eq: EH: u equation} and \eqref{eq: EH: w equation}. To that end, we have set the crossover scale such that $w(\tilde\chi)=0.5$ at $\tilde\chi=1$. The effective potential converges to the IR fixed point \eqref{eq: infrared fixed point} for $\tilde\chi\to \infty$. For $\tilde\chi\to 0$ one finds $u(\tilde\chi\to 0)\approx 0.0040$ and  $u(0)/w(0)^2=12.9$. Comparison with Figs.~\ref{fig: w} and \ref{fig: u} shows the same qualitative behavior, with only minor quantitative differences. We conclude that the influence of the couplings $C$ and $D$ on the scaling solutions for $u$ and $w$ is indeed small.

\subsection{$U+R+R^2$ system}
Next, we extend the system by including a term quadratic in the Ricci curvature scalar ($D=0$). This corresponds to the $f(R)$-type truncation in quadratic order. 
The flow equations in this system are given by
\al{
\p_\tau u(\tilde\chi)&=-4u(\tilde\chi)+\frac{1}{32\pi^2}M_U(\tilde\chi)\Big|_{d=0}\,,
\label{eq: flow equation of cosmological constant in URR2}
\\[1ex]
\p_\tau w(\tilde\chi)&=-2w(\tilde\chi) - \frac{1}{96\pi^2}M_F(\tilde\chi)\Big|_{d=0}\,,
\label{eq: flow equation of Planck mass in URR2}
\\[1ex]
\p_\tau C(\tilde\chi)&=-\frac{1}{576\pi^2}M_C(\tilde\chi)\Big|_{d=0}\,.
\label{eq: flow equation of C in URR2}
}
We find a UV asymptotically safe fixed point as
\al{
&u_*=0.000177\,,&
&w_*=0.0095\,,&
&C_*=-0.00292\,,
\label{eq: UV fixed point value of URR2}
}
while the flow equations \eqref{eq: flow equation of cosmological constant in URR2}--\eqref{eq: flow equation of C in URR2} do not admit the asymptotically free fixed point.
At the fixed point \eqref{eq: UV fixed point value of URR2}, the critical exponents are obtained to be
\al{
&\theta_{1,2}=2.48 + 0.341 i\,,&
&\theta_3=306.64\,.
}
These results are well compatible with the findings of other versions of the flow equations in the same truncation of the effective action. The scaling solutions for $u$ and $w$ look similar to the ones found for the flow from the AF-fixed point to the IR-fixed point.

\subsection{$U+R+C_{\mu\nu\rho\sigma}C^{\mu\nu\rho\sigma} $ system}
We finally investigate the system with $C=0$, retaining the effective potential, the curvature scalar $R$ and the squared Weyl tensor $C_{\mu\nu\rho\sigma}C^{\mu\nu\rho\sigma}$.
The beta functions read
\al{
\p_\tau u(\tilde\chi)&=-4u(\tilde\chi) +\frac{1}{32\pi^2}M_U(\tilde\chi)\Big|_{c=0}\,,\\[1ex]
\p_\tau w(\tilde\chi)&=-2w(\tilde\chi) - \frac{1}{96\pi^2}M_F(\tilde\chi)\Big|_{c=0}\,,
\label{eq: w in wD}
\\[1ex]
\p_\tau D(\tilde\chi)&= \frac{1}{960\pi^2}M_D(\tilde\chi)\Big|_{c=0}\,.
\label{eq: D in wD}
}

For pure gravity, we find the following fixed point value
\al{
&u_*=-0.00177\,,&
&w_*=0.0047\,,&
&D_*=-0.00185\,,
}
at which the critical exponents are
\al{
&\theta_1=3.96\,,&
&\theta_2=1.43\,,&
&\theta_3=-47.86\,.
}
Besides, the beta functions \eqref{eq: w in wD} and \eqref{eq: D in wD} admit the asymptotically free fixed point, i.e.
\al{
&u_*=0.00668\,,&
&w_*= 0.0132\,,&
&D_*^{-1}=0\,.
\label{eq: AFFP in URC2}
}
As expected, one finds the canonical scaling at this fixed point.
The IR fixed point \eqref{eq: infrared fixed point} is approached for $\tilde\chi\to \infty$.

\begin{figure*}
\includegraphics[width=\columnwidth]{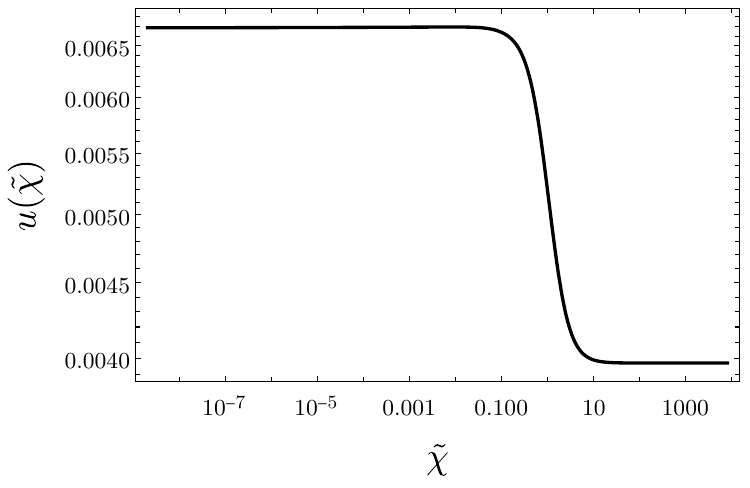}
\hspace{2ex}
\includegraphics[width=\columnwidth]{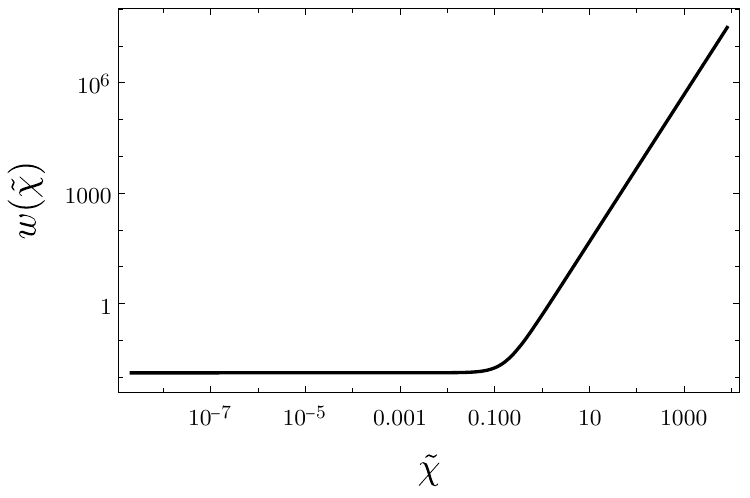}
\hspace{2ex}
\includegraphics[width=0.9\columnwidth]{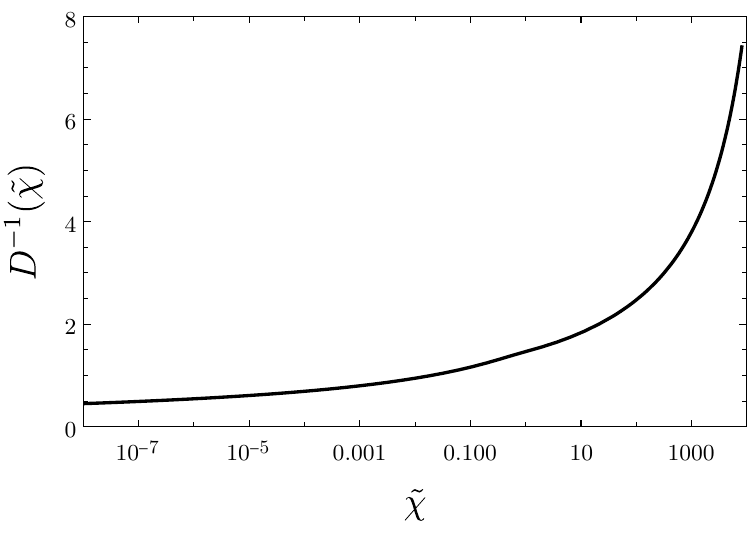}
\hspace{2ex}
\includegraphics[width=1.0\columnwidth]{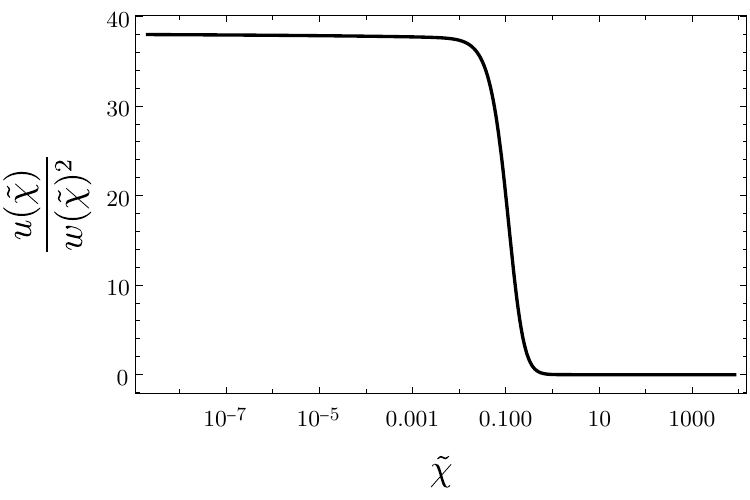}
\caption{
Critical trajectory of $u(\tilde\chi)$ (top-left), $w(\tilde\chi)$ (top-right), $D^{-1}(\tilde\chi)$ (bottom-left) and $u(\tilde\chi)/w(\tilde\chi)^2$ (bottom-right) as functions of $\tilde \chi$  in the $U+R+C^2$ truncation, for $N_S=N_V=N_F=0$. 
}
\label{fig: w and u in U+R+C2 truncation} 
\end{figure*}

Fig.~\ref{fig: w and u in U+R+C2 truncation} represents a critical flow trajectory from the asymptotically free fixed point \eqref{eq: AFFP in URC2} to the IR fixed point \eqref{eq: infrared fixed point}.
The critical flow trajectory is given such that at $\tilde\chi=1$ $w(\tilde\chi)=0.5$ and $D^{-1}(\tilde\chi=1)=1.47$. As for the other models one has to tune the value of $u(\tilde\chi=1)$ in order to realize the scaling solution, see Fig.~\ref{fig: u tuned}. This leads to $u(\tilde\chi\to 0)=0.0040$ and $u(0)/w(0)^2=38.3$. The scaling solutions for $u$ and $w$ are again very similar to Figs.~\ref{fig: w} and \ref{fig: u}. We have now selected a critical trajectory with a large value of $D^{-1}$ which grows for large $\tilde\chi$ outside the perturbative domain. For $\tilde\chi\to 0$ the asymptotically free fixed point $D^{-1}(\tilde\chi=0)=0$ is reached slowly. One also observes the logarithmic running of $D$ for the infrared regime $\tilde\chi\to \infty$.

\section{Conclusions}
\label{sec: Conclusions}
Within a general truncation of the functional flow equations for quantum gravity with up to four derivatives of the metric we have demonstrated the existence of a scaling solution or critical trajectory from the asymptotically free ultraviolet fixed point to the infrared fixed point. If this result remains valid beyond our truncation, quantum gravity can be formulated as a renormalizable asymptotically free quantum field theory for the metric coupled to other fields for particles. Asymptotic freedom constitutes a possible alternative to asymptotic safety. For certain models, as certain grand unified theories for the particle sector, no asymptotically safe ultraviolet fixed point may exist. In this case asymptotic freedom would remain as the only possibility. 

We have computed critical trajectories for a quantum field theory of the metric coupled to a scalar field $\chi$. Universal scaling functions for the dimensionless potential $u$, the coefficient of the curvature scalar $w$ or the coefficients of the four-derivative terms $C$ and $D$ depend on the dimensionless ratio $\tilde\chi=
\chi/k$. All our results transfer to scale dependent couplings for quantum gravity without the scalar field $\chi$. In this case one replaces $\chi^2$ by $2M_{\rm p}^2/\xi_{\infty}$, where $M_{\rm p}$ is the fixed Planck mass and the constant $\xi_\infty$ can be read off from our results in the limit $\tilde\chi\to \infty$
\al{
\xi_\infty =\lim_{\tilde\chi\to \infty} \left(\frac{4w}{\tilde\chi^2}\right)\,.
}
The scaling solution with increasing $\tilde\chi$ corresponds then to the general flow of scale dependent couplings with decreasing $k$. The fixed Planck mass $M_{\rm p}$ corresponds to a relevant parameter.
 
In our truncation we find whole families of scaling solutions. As always for a crossover between fixed points one parameter --- in our case typically $w$ --- only specifies at which value of $k$ the crossover from the UV to the IR takes place. In our approximation the family of trajectories is characterized by two more parameters that may be taken as $C(\bar \rho)$ and $D(\bar \rho)$ for some arbitrarily chosen $\tilde\rho=\bar\rho$. It seems likely, even though not fully established yet, that a trajectory exists which starts for $k\to \infty$ at the asymptotically free fixed point, and reaches for $k\to 0$ the asymptotically safe fixed point. This would imply that by tuning the parameters $C(\bar\rho)$ and $D(\bar\rho)$, which specify the members of the family of scaling solutions from the AF-fixed point to the IR-fixed point, one can obtain a specific trajectory which starts at the AF-fixed point in the ultraviolet, passes arbitrarily close, and therefore for an arbitrarily large interval of $\log k$, to the AS-fixed point, and finally ends in the IR-fixed point. This type of crossover involving three fixed points ``combines" features of all three fixed points. 

Along the critical trajectories the flow of $C$ and $D$, or $C^{-1}$ and $D^{-1}$, is found to be very slow. In practice, one can often approximate the functions $C(\tilde\rho)$ and $D(\tilde\rho)$ by constants $C(\bar \rho)$ and $D(\bar\rho)$ for an appropriate large range of $\tilde\rho$ around $\bar\rho$. Nevertheless, the logarithmic running of $C$ and $D$ continues for all values of $\tilde\rho=\chi^2/(2k^2)$. The coupling $D$ influences the propagator of the graviton. Our approximation directly yields the graviton propagator at zero momentum. For a generalization to nonzero squared momentum $q^2$ one may expect that $q^2$ replaces the infrared cutoff $k^2$ once $q^2$ exceeds $k^2$. In the infrared region with $w\gg 1$, $u/w\ll1$ this leads in Eq.~\eqref{eq: mass parameters} to $\tilde m_t^2\approx d$. Restoring dimensions and replacing $k^2\to k^2+q^2$ the inverse graviton propagator becomes
\al{
G_{\rm grav}^{-1} \sim \xi\chi^2 q^2 + D \left( \frac{\chi^2}{2(q^2+k^2)}\right) q^4\,, 
}
with effective Planck mass given by $\sqrt{\xi}\chi$. The replacement $D(\tilde\rho)\to D\left(\frac{\chi^2}{2(q^2+k^2)}\right)$ seems well justified in view of the mild logarithmic $\tilde\rho$-dependence of $D(\tilde\rho)$ and the fact that $q^2$ acts as an independent IR-cutoff. The momentum dependence of $D$ influences the poles of the graviton propagator and therefore the issue of a potential ghost- or tachyon-instability~\cite{Wetterich:2019qzx}. A detailed investigation will be necessary in order to see if the non-polynomial form of the inverse graviton propagator can cure the ``classical" instability of asymptotically free gravity.

For the scaling solution the dimensionless effective potential $u(\tilde\rho)$ takes a simple form. It interpolates smoothly between two constants for $\tilde\rho\to 0$ and $\tilde\rho\to \infty$. This form deviates strongly from a polynomial with a finite number of powers of $\tilde\rho$. The dimensionless curvature coefficient or squared effective Planck mass $w(\tilde\rho)$ is found to make a crossover from a constant value $w_0=w(\tilde\rho=0)$ to an increase linear in $\tilde\rho$ for large $\tilde\rho$. The value of $\tilde\rho$ where this change  takes place characterizes the location of the crossover. We have found this characteristic behavior for $u(\tilde\rho)$ and $w(\tilde\rho)$ in all truncations. This qualitative feature seems to be a rather robust result. 

The qualitative features of the crossover solution can be summarized by the approximation 
\al{
&w(\tilde\rho) = w_0 + \xi \tilde\rho\,,\nn[1ex]
&u(\tilde\rho) = u_\infty+ (u_0 -u_{\infty}) t_g(\tilde\rho, \bar\rho)\,,
}
where $t_g(\tilde\rho,\bar\rho)$ is a smooth ``threshold function" with $t_g(\tilde\rho \gg \bar\rho)=0$, $t_g(\tilde\rho\ll \bar\rho)=1$ and $\bar\rho=w_0/\xi$. The potential in the Einstein frame for the metric is given by
\al{
U_E= \frac{u(\tilde\rho)M_{\rm p}^4}{4w^2(\tilde\rho)}\,,
}
with $M_{\rm p}$ the fixed Planck mass introduced by  the Weyl scaling to the Einstein frame. We observe a flat plateau for $\tilde\rho\to 0$ and a decrease $\sim \tilde\rho^{-2}$ for $\tilde\rho\to\infty$,
\al{
\frac{U_E}{M_{\rm p}^4} =\frac{u_\infty + (u_0 -u_{\infty}) t_g(\tilde\rho, \frac{w_0}{\xi})}{4(w_0+\xi \tilde\rho)^2}\,.
\label{eq: ration UE M4}
}
This behavior is clearly visible in Figs.~\ref{fig: u/w2}, \ref{fig: w and u and C and D comp}, \ref{fig: w and u in EH truncation} and \ref{fig: w and u in U+R+C2 truncation}. For any cosmology for which $\tilde\rho$ diverges as time increases to infinity the cosmological constant problem is solved dynamically. For positive $u_0$ the flat tail of the potential for $\tilde\rho\to 0$ can describe some type of inflationary epoch. The form \eqref{eq: ration UE M4} is certainly an oversimplification. Nevertheless it is encouraging that already rather simple truncations of flow equations seem to imply interesting characteristic features of cosmology.

\subsubsection*{Acknowledgements}
This work is supported by the DFG Collaborative Research Centre ``SFB 1225 (ISOQUANT)" and Germany’s
Excellence Strategy EXC-2181/1-390900948 (the Heidelberg Excellence Cluster STRUCTURES).
M.\,Y. would like to thank the Yukawa Institute for Theoretical Physics at Kyoto University for support and hospitality by the long term visitor program of FY2022.

\onecolumngrid
\appendix
\section{Flow kernels}
\label{app: flow kernels}
In this appendix, we list the explicit forms of the flow kernels $M_i$ arising from the metric fluctuations, as computed in Ref.~\cite{Sen:2021ffc}.
\al{
&M_U=\frac{20}{3}(3d+2)\ell_{0}^4(\tilde m_t^2) + \frac{3}{10} (24 +80c -5v) \ell_0^4(\tilde m_\sigma^2)\,,
\label{eq: MUTT}
\\[2ex]
&M_F=-\bigg[ \frac{5}{2}(4d+3) \ell_0^2(\tilde m_t^2)
+\frac{40}{3}(3d+2) \ell_1^4(\tilde m_t^2)
+\frac{15}{2}(2c+d)(5+8d)\ell_1^6(\tilde m_t^2) \bigg] \nn
&\qquad\qquad
+  \frac{1}{12}(144c-10v+45)\ell_0^2(\tilde m_\sigma^2)
-\frac{v}{20}(120c-7v+35)\ell_1^6(\tilde m_\sigma^2)
+\frac{1}{7}(126c-7v+36)\ell_1^8(\tilde m_\sigma^2)\nn
&\qquad\qquad
+\frac{27c}{7}(224c-12v+63)\ell_1^{10}(\tilde m_\sigma^2)
-\frac{5}{3}\ell_0^2(0)
\label{eq: MFTT}
\\[2ex]
&M_C=
-\frac{770}{9}(d+1)\ell_0^0(\tilde m_t^2)
+\frac{410}{9} (4 d+3) \ell_1^2(\tilde m_t^2) 
+\frac{40}{27} (3 d+2) (3 c+7 d) \ell_1^4(\tilde m_t^2)
+\frac{1120}{9} (3 d+2) \ell_2^4(\tilde m_t^2) \nn
&\qquad\qquad
+\frac{80}{3} (8 d+5) (9 c+5 d) \ell^6_2(\tilde m_t^2)
+432 (5 d+3) \left(2 c^2+2 c d+d^2\right) \ell^8_2(\tilde m_t^2)\nn
&\qquad\qquad
+(12 c-v+4) \ell_0^0 (\tilde m_\sigma^2)
-\frac{7}{20} v (80 c-5 v+24) \ell_1^4 (\tilde m_\sigma^2)
+\frac{11}{15} (120 c-7 v+35) \ell_1^6 (\tilde m_\sigma^2)
\nn
&\qquad\qquad
-\frac{8}{63} (183 c-10 d) (126 c-7 v+36) \ell_1^8 (\tilde m_\sigma^2)
+\frac{20}{21} (21 c-d) (126 c-7 v+36) \ell_1^8 (\tilde m_\sigma^2)
 \nn
&\qquad\qquad
+\frac{5}{63} v^2 (126 c-7 v+36) \ell_2^8 (\tilde m_\sigma^2)
-\frac{2}{7} v (224 c-12 v+63) \ell_2^{10} (\tilde m_\sigma^2)
+\frac{25}{36} (288 c-15 v+80) \ell_2^{12} (\tilde m_\sigma^2) \nn
&\qquad\qquad
-\frac{40}{3} c v (288 c-15 v+80) \ell_2^{12} (\tilde m_\sigma^2)
+\frac{80}{3} c (1080 c-55 v+297) \ell_2^{14} (\tilde m_\sigma^2) 
+\frac{60480}{11} c^2 (220 c-11 v+60) \ell_2^{16} (\tilde m_\sigma^2) \nn
&\qquad\qquad
+\frac{5 d+3}{50(1+\tilde m_\sigma^2)}\ell_1^8(\tilde m_t^2)
+\frac{d (12 d+7) }{15 (1+\tilde m_\sigma^2)}\ell_1^{10}(\tilde m_t^2)
-\frac{400 (7 d+4) (d-6 c)^2}{21 (1+\tilde m_\sigma^2)} \ell_1^{12}(\tilde m_t^2)\,,
\label{eq: MCTT}
\\[2ex]
&M_D=
\frac{1030}{9} (d+1) \ell_0^0(\tilde m_t^2)
+\frac{500}{9} (4 d+3) \ell_1^2(\tilde m_t^2)
+\frac{200}{27} (3 d+2) (6 c-13 d) \ell_1^4(\tilde m_t^2)
+\frac{2800}{9} (3 d+2) \ell_2^4 (\tilde m_t^2)\nn
&\qquad\qquad
+\frac{1600}{3} d (8 d+5) \ell_2^6 (\tilde m_t^2)
+4080 d^2 (5 d+3) \ell_2^8 (\tilde m_t^2)\nn
&\qquad\qquad
+\frac{(126 c-7 v+36) }{210 (1+\tilde m_t^2)}\ell_1^8(\tilde m_\sigma^2) 
+\frac{(126 c-7 v+36) }{210 (1+\tilde m_t^2)}\ell_1^8 (\tilde m_\sigma^2)
+\frac{3 d (224 c-12 v+63) }{140 (1+\tilde m_t^2)}\ell_1^{10}(\tilde m_\sigma^2) \nn
&\qquad\qquad
-\frac{25 (d-6 c)^2 (288 c-15 v+80) }{9 (1+\tilde m_t^2)}\ell_1^{12} (\tilde m_\sigma^2)
-\frac{17}{18}\ell_0^0(0) \nn
&\qquad\qquad
+(12 c-v+4) \ell_0^0 (\tilde m_\sigma^2)
+\frac{5}{4} v (-80 c+5 v-24) \ell_1^4(\tilde m_\sigma^2) 
+ \frac{13}{3} (120 c-7 v+35) \ell_1^6(\tilde m_\sigma^2)
 \nn
&\qquad\qquad
-\frac{20}{63} (183 c-10 d) (126 c-7 v+36) \ell_1^8 (\tilde m_\sigma^2)
+\frac{5}{14} v (224 c-12 v+63) \ell_2^{10} (\tilde m_\sigma^2)
+\frac{25}{36} (288 c-15 v+80) \ell_2^{12}(\tilde m_\sigma^2)\nn
&\qquad\qquad
+\frac{5}{63} v^2 (126 c-7 v+36) \ell_2^8(\tilde m_\sigma^2) 
+\frac{50}{3} c v (288 c-15 v+80) \ell_2^{12}(\tilde m_\sigma^2) 
+\frac{80}{3} c (1080 c-55 v+297) \ell_2^{14}(\tilde m_\sigma^2)\nn
&\qquad\qquad
+\frac{60480}{11} c^2 (220 c-11 v+60) \ell_2^{16}(\tilde m_\sigma^2)  \nn
&\qquad\qquad
+\frac{5 d+3}{5(1+  \tilde m_\sigma^2)} \ell_1^8 (\tilde m_t^2)
+\frac{2d (12 d+7) \ell_1^{10} (\tilde m_t^2)}{3(1+  \tilde m_\sigma^2)}
-\frac{4000 (7 d+4) (d-6 c)^2 }{21(1+ \tilde m_\sigma^2)}\ell_1^{12} (\tilde m_t^2)\nn
&\qquad\qquad
+\frac{(126 c-7 v+36) }{21 (1+\tilde m_t^2)}\ell_1^8 (\tilde m_\sigma^2)
+\frac{3 d (224 c-12 v+63) }{14 (1+\tilde m_t^2)}\ell_1^{10} (\tilde m_\sigma^2) 
-\frac{250 (d-6 c)^2 (288 c-15 v+80) }{9 (1+\tilde m_t^2)}\ell_1^{12} (\tilde m_\sigma^2)\nn
&\qquad\qquad
-\frac{341}{18}\ell_0^0(0)\,.
\label{eq: MDTT}
}
Here the general threshold functions are defined by
\al{
\ell_p^{2n} (x) = \frac{1}{n!}\frac{1}{(1 + x)^{p+1}}\,.
}

\section{Gauss-Bonnet term}
\label{app: sec: Gauss-Bonnet term}
In this appendix we supplement a Gauss-Bonnet like term in our truncation \eqref{Eq: effective action for gravity part} of the effective average action 
\al{
\Gamma_{\rm GB} =\int_x\sqrt{g} \,E\left(R^2-4R_{\mu\nu}R^{\mu\nu}+R_{\mu\nu\rho\sigma}R^{\mu\nu\rho\sigma}\right)\,.
\label{eq: app: Gauss-Bonnet term}
}
The inclusion of this term completes the most general form with up to four derivatives of the metric. For constant $E$ Eq.~\eqref{eq: app: Gauss-Bonnet term} is a topological invariant. As a consequence, the function $E(\tilde\rho)$ does not appear in the flow equation for the other couplings. In turn, the flow equation for $E$ reads
\al{
(\p_t - 2\tilde\rho\,\p_{\tilde\rho}) E = -\frac{1}{5760\pi^2}\left( N_S+62N_V+\frac{11}{2}N_F +M_E \right)\,.
} 
The flow generator $M_E$ is given by
\al{
&M_E=
\frac{430}{3} (d+1) \ell_0^0 (\tilde m_t^2) 
+\frac{500}{3} (4 d+3) \ell_1^2 (\tilde m_t^2)
+\frac{200}{9} (3 d+2) (6 c+5 d) \ell_1^4 (\tilde m_t^2)
+\frac{1600}{3} (3 d+2) \ell_2^4 (\tilde m_t^2) \nn
&\qquad\qquad
+400 d (8 d+5) \ell_2^6 (\tilde m_t^2)
+6480 d^2 (5 d+3) \ell_2^8 (\tilde m_t^2)\nn
&\qquad\qquad
+(12 c-v+4) \ell_0^0(\tilde m_\sigma^2)
+\frac{15}{4} v (-80 c+5 v-24) \ell_1^4 (\tilde m_\sigma^2)
+13 (120 c-7 v+35) \ell_1^6 (\tilde m_\sigma^2) \nn
&\qquad\qquad
-\frac{20}{21} (183 c-10 d) (126 c-7 v+36) \ell_1^8(\tilde m_\sigma^2)
+\frac{15}{14} v (224 c-12 v+63) \ell_2^{10} (\tilde m_\sigma^2)
+\frac{25}{12} (288 c-15 v+80) \ell_2^{12} (\tilde m_\sigma^2) \nn
&\qquad\qquad
+\frac{5}{21} v^2 (126 c-7 v+36) \ell_2^8 (\tilde m_\sigma^2)
+50 c v (288 c-15 v+80) \ell_2^{12} (\tilde m_\sigma^2)
+80 c (1080 c-55 v+297) \ell_2^{14} (\tilde m_\sigma^2)\nn
&\qquad\qquad
+\frac{181440}{11} c^2 (220 c-11 v+60) \ell_2^{16} (\tilde m_\sigma^2)\nn
&\qquad\qquad
+\frac{3 (5 d+3) }{5(1+ \tilde m_\sigma^2)}\ell_1^8 (\tilde m_t^2)
+\frac{2 d (12 d+7) }{1+ \tilde m_\sigma^2}\ell_1^{10} (\tilde m_t^2)
-\frac{4000 (7 d+4) (d-6 c)^2 }{7(1+ \tilde m_\sigma^2)}\ell_1^{12} (\tilde m_t^2)
-360 (d+1) \frac{N}{\chi_E} \ell_0^0 (\tilde m_t^2)\nn
&\qquad\qquad
+\frac{(126 c-7 v+36) }{7 (1+\tilde m_t^2)}\ell_1^8(\tilde m_\sigma^2)
+\frac{9 d (224 c-12 v+63) }{14 (1+\tilde m_t^2)}\ell_1^{10}(\tilde m_\sigma^2)
-\frac{250 (d-6 c)^2 (288 c-15 v+80) }{3 (1+\tilde m_t^2)}\ell_1^{12}(\tilde m_\sigma^2)\nn
&\qquad\qquad
-\frac{317}{6}\ell_0^0(0) \,.
\label{eq: METT}
}

\bibliographystyle{JHEP} 
\bibliography{refs.bib}
\end{document}